\documentclass[aps,prl,reprint,showpacs,noshowkeys,floatfix]{revtex4-1}

\usepackage{graphicx}

\usepackage{amsmath}
\usepackage{mathtools}
\usepackage{amssymb}
\usepackage{amsfonts}
\usepackage{mathrsfs}
\usepackage[normalem]{ulem}
\usepackage{textcase}

\usepackage[backref=none]{hyperref}
\usepackage[usenames,dvipsnames]{color}
\definecolor{MyDarkBlue}{rgb}{0,0.1,0.7}
\hypersetup{pdfborder={0 0 0},colorlinks,breaklinks=true,
  urlcolor={MyDarkBlue},citecolor={MyDarkBlue},linkcolor={MyDarkBlue}
}

\newcommand{\rmi}{\mathrm{i}}
\newcommand{\rme}{\mathrm{e}}
\newcommand{\rmd}{\mathrm{d}}

\newcommand{\bq}{{\bf q}}

\newcommand{\nb}{{\nu}}
\newcommand{\bqi}{{\bf q}^{\rm i}}
\newcommand{\bqf}{{\bf q}^{\rm f}}

\newcommand{\lamT}{\lambda_{\rm T}}
\newcommand{\kB}{k_{\rm B}}
\newcommand{\eff}{\mathrm{eff}}
\newcommand{\Vext}{U}
\newcommand{\rhoeff}{{\rho_0}}
\newcommand{\asc}{{\alpha^{\rm sc}}}
\newcommand{\esc}{{E_0^{\rm sc}}}
\newcommand{\esccrit}{{E_{0,{\rm crit}}^{\rm sc}}}
\newcommand{\Dsc}{{\Delta_\infty^{\rm sc}}}
\newcommand{\subnn}{{n_1, n_2}}
\newcommand{\subnnm}{{n_1, n - n_1}}
\newcommand{\subuu}{{1,1}}
\newcommand{\suppm}{{\pm}}
\newcommand{\supp}{{+}}
\newcommand{\supm}{{-}}

\DeclareMathOperator*{\Tr }{Tr}

\DeclareMathOperator*{\Linv}{\mathscr{L}^{-1}_\beta\!}
\DeclareMathOperator*{\Linvs}{\mathscr{L}^{-1}_s\!}
\DeclareMathOperator*{\Linvab}{\mathscr{L}^{-1}_ {a \beta}\!}
\DeclareMathOperator*{\atan}{{\rm tan}^{-1}}
\DeclareMathOperator*{\erfc}{{\rm erfc}}
\newcommand{\Li}{\operatorname{Li}} 

\newcommand{\eqTL}{\stackrel{\rm TL}{=}}
\newcommand{\simTL}{\stackrel{\rm TL}{\simeq}}

\newcommand{\ie}{\textrm{i.e.}}
\newcommand{\eg}{\textrm{e.g.}}

\newcommand{\wrt}{\textrm{w.r.t.}\ }

\newcommand{\eref}[1]{(\ref{#1})}
\newcommand{\Eref}[1]{Equation~(\ref{#1})}
\newcommand{\Leref}[1]{Eq.~(\ref{#1}) of the main text}

\newcommand{\fref}[1]{Fig.~\ref{#1}}
\newcommand{\Lfref}[1]{Fig.~\ref{#1} of the main text}

\newcommand{\Fref}[1]{Figure~\ref{#1}}

\begin{document}

\title{
Partial Fermionization---Spectral Universality in 1D Repulsive Bose Gases
}
\date{\today}

\author{Quirin Hummel}
\email{quirin.hummel@ur.de}
\author{Juan Diego Urbina}
\author{Klaus Richter}
\affiliation{Institut f\"ur Theoretische Physik, Universit\"at Regensburg, D-93040
Regensburg, Germany}


\begin{abstract}
Due to the vast growth of the  many-body level density with excitation energy, its smoothed form is of central relevance for spectral and thermodynamic properties of interacting quantum systems.
We compute the cumulative of this level density for confined one-dimensional continuous systems with repulsive short-range interactions.
We show that the crossover from an ideal Bose gas to the strongly correlated, fermionized gas, \ie,  partial fermionization, exhibits universal behaviour:
Systems with very few up to many particles share the same underlying spectral features.
In our derivation we supplement quantum cluster expansions with short-time dynamical information.
Our nonperturbative analytical results are in excellent agreement with numerics for systems of experimental relevance in cold atom physics, such as interacting bosons on a ring (Lieb-Liniger model) or subject to harmonic confinement.
Our method provides predictions for excitation spectra that enable access to finite-temperature thermodynamics in large parameter ranges.
\end{abstract}



\maketitle


The huge progress in cold atom physics has enabled precision experiments which allow to confine, control and study ensembles of 
atoms with particle numbers ranging from very few \cite{Greiner2002,Serwane2011,Zuern2012,Wenz2013,Preiss2015} to nearly macroscopically many \cite{Moritz2003,Kinoshita2004,Paredes2004,Turpin2015}.
The high control over parameters, trapping to low dimensions and tunability of interactions has lead to a synergetic understanding of highly correlated many-body (MB) systems \cite{Bloch2008}, in many cases based on theories of one-dimensional integrable models \cite{Korepin1993,Guan2013} and correspondingly tailored experiments \cite{Kinoshita2006,Meinert2015}.
However, in situations deviating from integrability (see, \eg, \cite{Polkovnikov2011,Schmelcher2012,Wilson2014,Brandino2015,GarciaMarch2018}) the theoretical treatment of systems with an intermediate number $N$ of interacting identical particles is particularly hard, especially when the observed spectral, thermodynamic or dynamical properties involve highly excited multi-particle states.

The conceptual challenges are numerous:
First, systems with fixed $N$ require a canonical treatment, in particular when approaching the few-body regime, where grand canonical approaches often fail \cite{Kocharovsky2006}.
Second, due to strong inter-particle correlations that can experimentally be pushed up to the limit of fermionization in Bose gases \cite{Kinoshita2004,Paredes2004,Zuern2012,Guarrera2012,Ronzheimer2013}, and especially for small $N$, mean-field approaches or more generally $1/N$ expansions get problematic.
Elaborate MB techniques allow for calculating ground and low excited states of such interacting multi-particle systems with high precision.
However, these methods reach their limits when increasing $N$ or the degree of excitation since this implies vastly growing Hilbert space dimensions.

\begin{figure}[ttt]  
	\includegraphics[width=0.9\columnwidth]{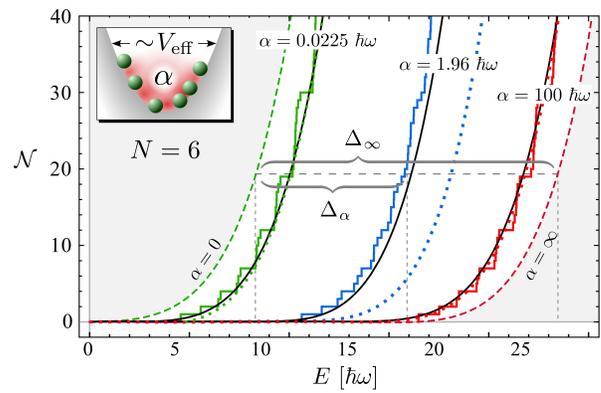}
	\caption{\label{fig:1DHON6}
		Many-body level counting function for six interacting bosons in a harmonic trap (spacing $\hbar \omega$) for different contact interaction strengths $\alpha$. Numerically exact results for ${\cal N}(E)$ (staircases) exhibit characteristic shifts $\Delta_\alpha$ in $E$ towards the limit $\alpha\!\rightarrow\!\infty$ of full fermionization.
		These shifts carry universal features and are quantitatively explained by our theory (solid lines) based on \eref{eq:EshiftChi},
		\eref{eq:xequ}. 
		Dotted lines denote analytical QCE-based approximations~\eref{eq:scalingNgen} invoking the limiting cases of weak and strong $\alpha$,
		see main text.
	}
\end{figure}

This goes along with a close to exponential increase of the MB density of states (DOS) with excitation energy for continuous
$N$-particle systems, even in the 1D case.
The universal Bethe law \cite{Bethe1936,Ramanujan1918} and variants \cite{vonEgidy1986, vonEgidy1988} for sufficiently lowlying excitations in large-$N$ fermionic systems represent a famous example in nuclear physics.
There, the effect of (residual) interactions is merely a broadening of the otherwise highly degenerate noninteracting MB spectrum \cite{French1970,Bohigas1971}, while for small to intermediate $N$ interactions have nontrivial effects and the Bethe law generally fails~\cite{Hummel2014,Andrews2003}.

Nonetheless, the spacing between MB levels and the associated fluctuations tend to zero such that individual highly-excited MB levels are usually no longer resolvable.
Hence the (locally) energy-averaged, smooth MB DOS $\rho^{(N)}(E)$ gains particular relevance
\footnote{%
	See, \eg, Refs.~\cite{GarciaGarcia2017} and \cite{Bogomolny2014} for recent calculations for the Sachdev-Ye-Kitaev model and quantum spin-1/2 Ising model with tilted field.
}.
In particular it plays the central role for computing thermodynamic equilibrium properties at finite temperature.
Beyond that, $\rho^{(N)}(E)$ is a key ingredient to nonequilibrium quantum work statistics that has drawn much attention lately \cite{Campisi2011,Goold2018,Wang2017,Arrais2018}, not least due to a recently revealed connection to information scrambling \cite{Chenu2018}.

This calls for developing genuinely interacting MB techniques specifically devised to directly compute the smooth DOS, thereby circumventing the intricate or simply impossible calculation of individual (highly) excited MB levels which requires additional information that is afterwards 
smoothed out anyway.

Similar to the single particle case \cite{Weyl1911,Balian1970,Balian1971,Ivrii2016}, a smooth MB DOS corresponds to and requires dynamical information from MB quantum propagation on finite time scales only.
Invoking such short-time information in a quantum cluster expansion (QCE) \cite{Ursell1927,Kahn1938,Grueter1995I} implies, as we will show, that interaction effects in the smooth DOS arise nonperturbatively from universal cluster kernels dressed with terms depending on the confinement potential.
Specifically, we consider not directly the MB DOS $\rho^{(N)}(E)$ but the (smooth) MB level counting function ${\cal N}(E) \!=\! \int_0^E {\rm d}E' \rho^{(N)}(E')$, depicted in Fig.~\ref{fig:1DHON6} for a harmonically trapped Bose gas. 
${\cal N}(E)$ exhibits interaction-dependent characteristic horizontal shifts $\Delta_\alpha$ indicating what we call {\em partial fermionization}.  
We will analytically show that these shifts, and thereby ${\cal N}$ and $\rho^{(N)}$, follow with high accuracy $N$-independent universal laws, \ie, broad classes of interacting bosonic systems ranging from very few to many particles possess equal spectral features.
Remarkably, these robust features are reminiscent of the spectral shifts in the famous solvable Calogero-Sutherland models \cite{Calogero1971III,Sutherland1972bII} which admit an interpretation in terms of fractional exclusion statistics \cite{Haldane1991,Murthy1994,Ha1994}.

We first outline the main steps of our QCE for the canonical partition function providing the basis for our further (asymptotic) analysis to derive our main result, a universal law for partial fermionization.


\textit{Canonical partition function}.---%
The MB DOS $\rho^{(N)}_{\pm}(E)$ of a system of $N$ identical quantum particles (``$\pm$'' denoting bosons and fermions) is related to the canonical partition function $Z^{(N)}_{\pm}(\beta)$ through the inverse Laplace transform
$\rho^{(N)}_{\pm}(E) \!=\!  \Linv[Z^{(N)}_{\pm}(\beta)](E)$ with $\beta\!=\!1/(k_{\rm B} T)$.
Furthermore, $Z^{(N)}_{\pm}(\beta)\!=\!\Tr_{\pm} K^{(N)}(t \!=\! -\rmi \hbar \beta)$ is the trace over the propagator $K^{(N)}$ for $N$ {\it distinguishable} particles in the properly (anti-)symmetrized basis.

For $N$ {\em noninteracting} particles of mass $m$, each with coordinates ${\bf q}$, confined by a homogeneous potential $\Vext({\bf q})=w^{\mu}\Vext({\bf q}/w)$, it can be expressed in closed form \cite{Hummel2014} (see \hyperref[sec:sm:nonint]{Appendix A}),
\begin{equation} \label{eq:Znonint}
 Z^{(N)}_{0,\pm}(\beta) = \sum_{l=1}^N z^{(N,d)}_{\pm,l} \left( \frac{V_\eff}{\lambda_T^d} \right)^l \,,
\end{equation}
with universal constants $z^{(N,d)}_{\pm,l}$, physical dimension $D$ and effective dimension $d=D(1+\frac{2}{\mu})$.
Setting $\hbar^2/(2m)=1$, the thermal wavelength is $\lambda_T \!=\! \sqrt{4 \pi \beta}$ and the effective volume is $V_\eff = (4\pi)^{D / \mu} \int \rmd^D\!q \, \exp[{-\Vext({\bf q})}]$.
The case without external potential is included as $\mu \rightarrow \infty$, then $d=D$ and $V_\eff$ equals the physical volume $V_D$.

\begin{figure}[ttt]
		\includegraphics[width=0.95\columnwidth]{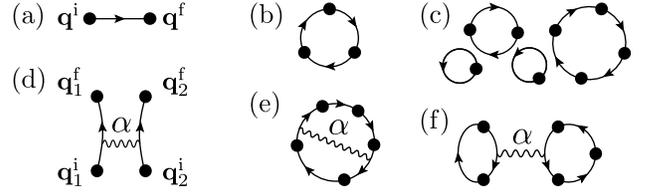}
	\caption{\label{fig:diagrams}
	Leading-order contributions to the quantum cluster expansion.
	(a) SP propagator $K_0^{(1)}(q^{\rm f}, q^{\rm i};t)$;
	(b) contribution $\mathcal{A}_{n}$ from a single cycle (here $n\!=\!3$);
	(c) specific clustering yielding $\mathcal{A}_{\mathfrak{N}}$ (here $\mathfrak{N}\!=\!\{1,1,2,4\}, N\!=\!8$);
	(d) interacting part $\Delta K^{(2)}_\alpha((q_1^{\mathrm{f}}, q_2^{\mathrm{f}}),(q_1^{\mathrm{i}},q_2^{\mathrm{i}});t)$ of the two-body propagator; 
	(e),(f) examples for intra-/inter-cycle contributions $\mathcal{A}_{\subnn}^{\mathrm{intra}}$ and $\mathcal{A}_{\subnn}^{\mathrm{inter}}$ with $n_1\!=\!3, n_2\!=\!2$.
	A diagram is given by the product of all SP and interacting two-body components $K_0^{(1)}$ and $\Delta K^{(2)}_\alpha$ after spatial integration over all (internal) points.
	}
\end{figure}

\textit{Quantum cluster expansion}.---%
The noninteracting part $K^{(N)}_0$ of the propagator factorizes into single-particle (SP) propagators, see \fref{fig:diagrams}(a).
A contribution to $Z^{(N)}_{0,\pm}$ corresponding to a permutation $P$ is a product of \textit{cluster} terms, resembling the decomposition of $P$ into cycles \cite{Rotman1999}.
Using the semigroup property of the SP propagator and identifying ${\bf q}_{n+1}\equiv {\bf q}_1\!=\! {\bf q}$, each cycle involving a subset of $n$ particles [see \fref{fig:diagrams}(b)] yields the amplitude $\mathcal{A}_{n}(t) \!=\! \int \rmd^D\!q \, K^{(1)}_0({\bf q}, {\bf q};n t)$.
In line with our major assumption of short-time propagation we can use \cite{Hummel2014} $K_0^{(1)}({\bf q},{\bf q};t) \simeq \exp [-\frac{\rmi}{\hbar}\Vext({\bf q})t] K_{\rm free}^{(1)}({\bf q},{\bf q};t)$ where $K_{\rm free}$ stands for unconfined propagation.
The full contribution to $Z_{0,\pm}^{(N)}$ of a permutation is then $\mathcal{A}_\mathfrak{N}(-\rmi \hbar \beta) \!=\! \prod_{n \in \mathfrak{N}} \mathcal{A}_{n}(- \rmi \hbar \beta)$, in terms of the multiset $\mathfrak{N} \!=\!  \{ n_1 , n_2 , \ldots , n_{|\mathfrak{N}|}\}$ of cycle lengths, see \fref{fig:diagrams}(c).
Further evaluation of these amplitudes eventually yields the explicit result \eref{eq:Znonint} (see \hyperref[sec:sm:nonint]{Appendix A} and \cite{Hummel2014}).

The implementation of {\em interaction effects} begins with a cluster expansion \cite{Ursell1927,Kahn1938,Grueter1995I} of $K^{(N)}$ to first order in
the interaction by decomposing the full two-body propagator $K^{(2)} = K^{(2)}_0 + \Delta K^{(2)}_{\alpha}$ into $K_0^{(2)}$ and nonperturbative interaction contributions $\Delta K^{(2)}_{\alpha}$ where $\alpha$ is an energy associated with the coupling strength
\footnote{%
	Setting $\hbar^2/(2m)\!=\!1$, any coupling of nonvanishing dimension $[q^\nu]$, $\nu \neq 0$, can be associated with an energy-like parameter $\alpha$, excluding only scale-invariant interactions.
}.
To calculate interaction effects we choose all pairs $\{k,l\}$ of particles and replace  
$K_0^{(1)}({\bf q}_{P(k)},{\bf q}_k;t) K_0^{(1)}({\bf q}_{P(l)},{\bf q}_l;t)$ in $\mathcal{A}_\mathfrak{N}$ by the interaction term 
$\Delta K^{(2)}_\alpha(({\bf q}_{P(k)},{\bf q}_{P(l)}),({\bf q}_k,{\bf q}_l);t)$, see \fref{fig:diagrams}(d).
The interaction can link two particles involved in either the same [see \fref{fig:diagrams}(e)] or in two different cycles [see
\fref{fig:diagrams}(f)] of $P$, referred to as {\it intra}- and {\it inter-cycle} contributions $\mathcal{A}^{\rm
intra/inter}_{\subnn}$
where $\subnn$ denotes the distribution of the $n = n_1 + n_2$ particles.
Evaluation of the diagram classes in Figs.~\ref{fig:diagrams}(e) and \ref{fig:diagrams}(f) yields
\begin{equation} \label{eq:DeltaZ}
	 Z^{(N)}_{\alpha,\pm} =Z^{(N)}_{0,\pm} +\sum_{n=2}^N (\pm 1)^n Z^{(N-n)}_{0,\pm} \sum_{n_1=1}^{n-1} \mathcal{A}_{\subnnm}^{\suppm}
\end{equation}
with amplitudes of the form
\begin{equation} \label{eq:Ascaling}
	\mathcal{A}_{\subnn}^{\suppm} = 
\frac{1}{2} [\mathcal{A}_{\subnn}^{\rm inter} \pm \mathcal{A}_{\subnn}^{\rm intra}]
=
\frac{V_\eff}{\lambda_T^d n^{d / 2}}  \, a_{\subnn}^{\suppm}(\beta \alpha) \,,
\end{equation}
definining the interaction kernels $ a_{\subnn}^{\suppm}(\beta \alpha)$, see below.

The philosophy behind cluster expansions implies that the form~\eref{eq:Ascaling} of amplitudes is generic for arbitrary short-range interactions \cite{Hummel2017thesis,HummelLong}:
The main contribution to the $n$-fold integrals involved stems from the region where all $n$ particles are close to each other, allowing us to extend all integrals over relative coordinates to infinity.
Hence the $a_{\subnn}^{\suppm}(\beta \alpha)$ do not depend on the external potential, a key feature of our approach that also generalizes to higher order, \ie, clusters involving $n$-body corrections.
Instead, only the center of mass is subject to $\Vext({\bf q})$, thus yielding the effective system size $V_\eff$
as prefactor in \eref{eq:Ascaling}.
In view of (\ref{eq:Znonint}--\ref{eq:Ascaling})
this yields the entire QCE partition function
\begin{equation} \label{eq:Z1}
	Z_{\alpha,\pm}^{(N)}(\beta)=\sum_{l=1}^{N} \left[ z_{\pm,l}^{(N,d)} + \Delta z_{\pm,l}^{(N,d)}(\beta \alpha) \right] \left( \frac{V_\eff}{\lambda_T^d} \right)^l
\end{equation}
with interaction-related terms $ \Delta z_{\pm,l}^{(N,d)}(\beta \alpha) $ given by the kernels, \ie, to first order, $ a_{\subnn}^{\suppm}(\beta \alpha) $ (see \hyperref[sec:sm:kernels]{Appendix B}).

Correspondingly, the general QCE expression for the central quantity $\mathcal{N}_\alpha(E) \!=\! \int_{0}^E \rmd E' \rho^{(N)}(E')$ is (see \hyperref[sec:sm:LCF]{Appendix C} and \cite{Hummel2017thesis}), to arbitrary order,
\begin{equation} \label{eq:scalingNgen}
	\mathcal{N}_\alpha(E) = \sum_{l=1}^N \! \left[ \frac{z^{(N,d)}_{\pm,l}}{\Gamma\left(\frac{ld}{2} \!+\! 1\right)} 
   \! + \! g_{\pm,l}^{(N,d)}\!\!\left( \frac{E}{\alpha} \right) \right]
 \!
		V_\eff^l \! \left( \frac{E}{4\pi} \right)^{ld/2}  \!\!\!\!\!\!.
\end{equation}
It features the same polynomial structure in $V_\eff E^{d/2}$ as its noninteracting counterpart, while the $g_{\pm,l}^{(N,d)}$ add a functional dependence on $E/\alpha$ to the coefficients given by the interaction kernels, \ie, to first order, $a_{\subnn}^{\suppm}(\beta \alpha)$.


\textit{Contact interaction}.---%
For explicit calculations
and motivated by the central importance for quasi 1D cold atom systems \cite{Bloch2008,Olshanii1998,Bergeman2003} we consider Hamiltonians
\begin{equation} \label{eq:Hdelta}
	\hat{H} =  \sum_{i=1}^N \left(-\frac{\partial^2}{\partial q_i^2} +\Vext(q_{i})\right)+ \sqrt{8 \alpha} \sum_{i<j} \delta(q_i - q_j)
\end{equation}
of $N$ interacting bosons with coordinates $q_i$ in 1D.
One obtains (see \hyperref[sec:sm:kernels]{Appendix B}) explicit analytical expressions for the kernels $a_{\subnn}^{\supp}(\beta \alpha)$ in~\eref{eq:Ascaling}.
Closed explicit expressions for the $g_{\pm,l}^{(N,d)}(E/\alpha)$ in (\ref{eq:scalingNgen}) follow 
for the prominent 1D cases of $\Vext(q)\!=\!0$ ($d\!=\!1$), harmonic confinement ($d\!=\!2$), and linear potential wells ($d\!=\!3$) (see \hyperref[sec:sm:LCF]{Appendix C}).

Before addressing representative cases we note that the QCE~(\ref{eq:Z1}, \ref{eq:scalingNgen}), evaluated to first-order, although devised for weak interaction, can also be applied to the complementary regime of strong coupling (see \hyperref[sec:sm:strong]{Appendix D}) by means of fermionization \cite{Tonks1936,Girardeau1960} due to an exact duality \cite{cheon1999} of strongly coupled bosons and weakly coupled spinless fermions.


\textit{Harmonic confinement}.---%
We first consider $\Vext(q_i) \!=\! (\hbar\omega)^2 q_i^2/4$, for which $V_\eff \!=\! 4\pi /(\hbar\omega)$,
and compare in \fref{fig:1DHON6} analytical QCE results (dotted lines) for ${\cal N}_\alpha(E)$ with extensive numerical calculations (staircases) based on exact diagonalization and hence restricted to roughly the first 40 excited MB levels for $N\!=\!6$. 
The first-order QCE, implemented as weak- and dual strong-coupling expansions, indeed is valid in the respective regimes.
However, for intermediate couplings (here $\alpha \simeq 2 \hbar \omega$) it degrades. 
Moreover, such deviations grow with increasing $N$ calling for an improved method that adequately treats intermediate couplings.


\textit{Partial fermionization}.---%
Interactions predominantly cause characteristic shifts of ${\cal N}_\alpha(E)$ towards larger energies (as visible in~\fref{fig:1DHON6}).
Presuming knowledge of the noninteracting spectra, the shifts $\Delta_\alpha$ of individual levels contain all information about the interacting spectra.
We adopt this reformulation of the problem to develop a method that directly addresses these shifts on average.
Our approach further enables asymptotic considerations that strongly simplify the MB problem and highlight the universality behind
partial fermionization.

For the interaction-induced energy shift at fixed  ${\cal N}$,
\begin{equation} \label{eq:EshiftDeltaDef}
\Delta_\alpha \equiv \langle E^{(n)}(\alpha) \! - \! E^{(n)}(0) \rangle_n \equiv \langle E^{(n)}(\alpha) \rangle_n - E_0   \, , 
\end{equation}
averaged over a bunch of individual MB levels $E^{(n)}$ we propose  the ansatz
\begin{equation} \label{eq:EshiftChi}
	\Delta_\alpha \approx \chi^{(N,d)}(E/\alpha) \, \Delta_\infty^{(N,d)}(E_0, V_\eff) \, , 
\end{equation}
where $E = E_0 + \Delta_\alpha$ is the shifted  energy, separating the  $V_\eff$-dependence from an $\alpha$-dependent function $\chi^{(N, d)}$, in view of the notable structure of $\mathcal{N}_\alpha$ within QCE~\eref{eq:scalingNgen} and corroborated by a general consistency argument (see \hyperref[sec:sm:consistency]{Appendix E, subsection 3} and \cite{Hummel2017thesis}).
$\Delta_\infty$ denotes the full ``horizontal'' shift (see \fref{fig:1DHON6}) between fermionized and noninteracting bosonic levels for fixed $\mathcal{N} \equiv \langle n \rangle_n = \mathcal{N}_0(E_0)$.
We find (see \hyperref[sec:sm:shifting1]{Appendix E})
\begin{equation} \label{eq:DeltaEinfty}
	\Delta_\infty^{(N,d)} \approx {\rm const.} \cdot  V_\eff^{-2/d} {\mathcal{N}}^{(2/d-1)/N} \,.
\end{equation}
The $\alpha$-dependent factor $\chi\! \in \! [0,1]$ in (\ref{eq:EshiftChi}) continuously interpolates between the free Bose gas $\chi \to 0$ and the fully fermionized gas $\chi \to 1$, quantifying \textit{partial fermionization}.
Most notably, the central function $\chi^{(N,d)}(E/\alpha)$ in \eref{eq:EshiftChi} is uniquely obtained from QCE \eref{eq:scalingNgen} by matching
\begin{equation} \label{eq:NequN}
	\mathcal{N}_\alpha(E) = \mathcal{N} =  \mathcal{N}_0(E_0) = \mathcal{N}_0(E - \Delta_\alpha)
\end{equation}
in the regime $E^{d/2} V_\eff \gg 1$ of weak quantum degeneracy, where the first-order QCE becomes 
increasingly accurate. 
For the LHS of \eref{eq:NequN} we apply QCE~\eref{eq:scalingNgen}, while for the RHS we use the result \eref{eq:Znonint} for  $\alpha\!=\!0$, and implement the shift $\Delta_\alpha$, Eq.~\eref{eq:EshiftChi}, as an expansion around $E$ in the small parameter $\Delta_\infty^{(N,d)} / E = \mathcal{O}(E^{-d/2} V_\eff^{-1})$.
Matching the next-to-leading order $\mathcal{O}(V_\eff^{N-1})$ in~\eref{eq:NequN} fixes $ \chi^{(N,d)}(E/\alpha) \propto - g^{(N,d)}_{+,N-1}(E/\alpha) $ (see \hyperref[sec:sm:shifting1]{Appendix E}), which, remarkably, is fully determined by two-body clusters for which the first-order QCE is exact.

A solution for ${\cal N}_\alpha(E)$ is achieved by determining the partial fermionization for a given initial \textit{noninteracting} energy $E_0$, reducing the problem to finding, in view of~\eref{eq:EshiftChi}, the root of
\begin{equation} \label{eq:xequ}
	x = \chi^{(N,d)}(E/\alpha) =  \chi^{(N,d)}((E_0 + x \, \Delta_\infty)/ \alpha) \, .
\end{equation}
This implicitly defines $x=\chi$ as a function of $E_0, \alpha$, and $N$.
The method efficiently emulates the effect of higher-order clusters in terms of the smallest ones, giving excellent predictions (see solid curves in \fref{fig:1DHON6}).
While the presented lowest-order version involves only two-body clusters it can be pushed to second (and higher) order in a controlled way \cite{Hummel2017thesis, HummelLong}, where three-body (and larger) clusters correct for multi-particle collision effects.

\begin{figure}[ttt]  
	\includegraphics[width=0.92\columnwidth]{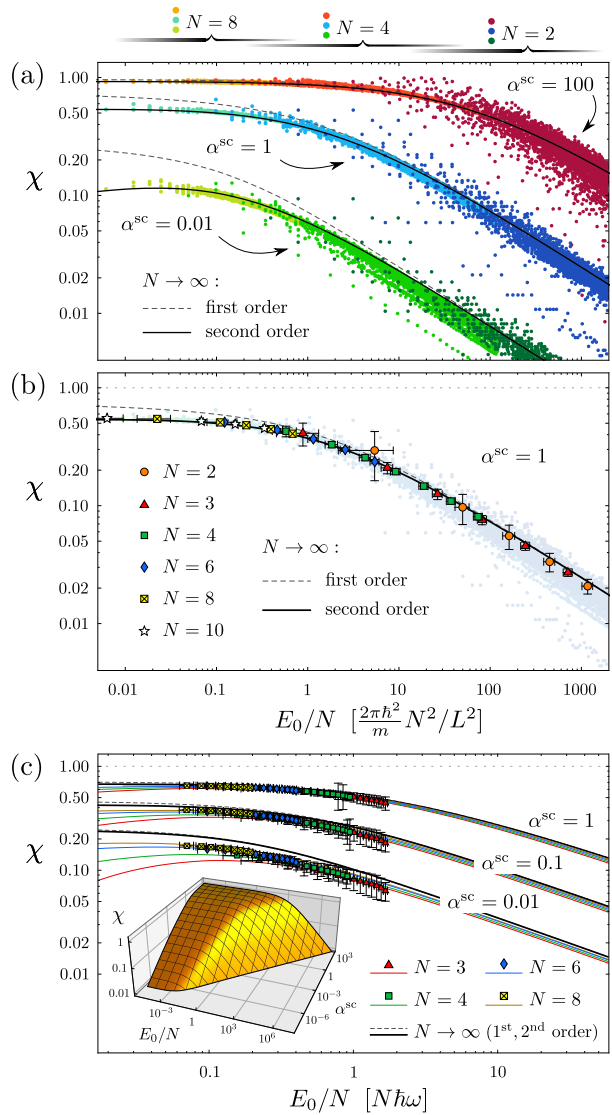}
	\caption{\label{fig:chi}
		Universal behavior of partial fermionization in repulsive 1D Bose gases.
		The ratio $\chi \! \in \! [0,1]$, Eq.~\eref{eq:EshiftChi}, as a function of energy per particle $E_0 / N$ of the noninteracting 
		system is shown for (a), (b) the Lieb-Liniger model and (c) harmonic confinement for various couplings $\alpha$ and $N$.
		Numerical results for $\chi$ [dots representing individual levels in (a); symbols in (b) with bars indicating one standard deviation of local average over data chunks to obtain smooth densities] are extracted level-by-level from comparing the staircase functions (see, e.g., \fref{fig:1DHON6}) of noninteracting, fully fermionized, and interacting energies [$\sim 10^6$ lowest states in (a), (b) and $\sim 80-200$ states in (c)].
		The numerical data for different $N$ collapse to universal functions (\ref{eq:universalchigen}), see text for details.
		Inset: universal prediction for $\chi(E_0^{\rm sc},\alpha^{\rm sc})$ for harmonic confinement.
	}
\end{figure}


\textit{Asymptotics and universality}.---%
An asymptotic analysis (see \hyperref[sec:sm:asymptotics]{Appendix F}) of $\chi$ and $\Delta_\infty$ for large $N$ further reveals that
\begin{equation} \label{eq:chi_infty}
x  \approx \lim_{N\rightarrow\infty}\chi^{(N,d)}(N \tilde\epsilon) = 1 - \rme^{d/(2 \tilde\epsilon)} \erfc{\sqrt{d/(2\tilde\epsilon)}} 
\end{equation}
where $\tilde\epsilon \!=\! [E_0^{\rm sc} + x \, \Delta_\infty^{\rm sc}(E_0^{\rm sc})]/\alpha^{\rm sc} $ and, with $n_\eff= N / V_\eff$,
\begin{equation} \label{eq:scaledvars}
	E_0^{\rm sc} = \frac{E_0}{N} \mathcal{E}^{-1} , \quad
		\alpha^{\rm sc} = \alpha \mathcal{E}^{-1} , \quad
		\mathcal{E} = \frac{2\pi \hbar^2}{m} n_\eff^{2/d} \,,
\end{equation}
implying a \textit{universal} law for the partial fermionization,
\begin{equation} \label{eq:universalchigen}
	\chi = \chi(d,E_0^{\rm sc}, \alpha^{\rm sc}) \, .
\end{equation}
For systems with different $N$ it predicts that $\chi$, and hence ${\cal N}_\alpha$, depend in the same peculiar way on $\alpha$ and the energy per particle $E_0/N$ of the corresponding noninteracting system, both appropriately scaled in terms of the energy unit $\mathcal{E}=(2\pi \hbar^2/m) n_\eff^{2/d}$, establishing a key feature of the observed universality:
It relates high excitations in large-$N$ systems to lowlying excitations in corresponding systems with smaller $N$.
Explicit approximants for $\chi(d,E_0^{\rm sc}, \alpha^{\rm sc})$ can be found by iteration (see \hyperref[sec:sm:regimes]{Appendix G}).

In Figs.~\ref{fig:chi}(a) and \ref{fig:chi}(b) we compare these predictions with numerically obtained data based on MB levels of~\eref{eq:Hdelta} for the paradigmatic Lieb-Liniger model \cite{Lieb1963I,Lieb1963II,McGuire1964,Jiang2015}  [$\Vext(q_i)\!=\!0$ on a ring with length $L$, \ie, $n_\eff \! = \! N/V_\eff \! = \! N/L$].
We find that the universality is fulfilled with remarkable accuracy for the whole range of interactions and particle numbers, even down to $N \!=\! 2$.
Moreover, for growing $N$ spectral fluctuations, not included in our analytical approach, are strongly suppressed, implying approximate analytical predictability of individual excited MB energies for arbitrary parameters.

Inaccuracies at very low energies and couplings are cured by extending the energy shifting from first order (dashed), based on two-body processes, to second order (solid) involving three-cluster diagrams which, again, can be calculated analytically \cite{Hummel2017thesis, Geiger2018}.
Our approach amounts to a description of the entire smooth spectrum in terms of only two- or three-body processes which nonperturbatively interpolates between $\alpha\!=\!0$ and $\alpha\!\rightarrow\!\infty$.

Figure \ref{fig:chi}(c) shows results for harmonic confinement, for which $\Delta_\infty^{\rm sc} \!=\! 1/2$, representing a generic nonintegrable $N$-particle system, see also \fref{fig:1DHON6}.
The full lines for $N\!=\! 3,4,6,8$ display the respective solutions of~\eref{eq:xequ} converging to the large-$N$ limit.
The universal prediction (inset) shows, besides fermionization $\chi\!\simeq\!1$ (roof) and the perturbative regime $\chi \simeq 2 \sqrt{ \alpha^{\rm sc} / (\pi E_0^{\rm sc})}$ (right flank), a nonperturbative quantum regime for $E_0^{\rm sc}\!\ll\!({\alpha^{\rm sc}})^{1/3}, \alpha^{\rm sc} \ll 1$ where $\chi \simeq 2 (\alpha^{\rm sc}/\pi)^{1/3}$ becomes independent of $E_0^{\rm sc}$ (see \hyperref[sec:sm:regimes]{Appendix G}).
This peculiarity connects our findings to the solvable Calogero-Sutherland model with harmonic trapping \cite{Calogero1971III,Sutherland1971aI} where exact spectra are subject to a constant shift that we identify as a specific realization of $\chi$:
Within our scheme, the dimensionless coupling constant there prohibits an explicit energy dependence of $\chi$ in~\eref{eq:xequ}.
Here we find a generalization (including nonintegrable systems) where $\chi$ is allowed to vary over the (smoothed) spectrum in a way characteristic for the particular type of interaction.
We stress that due to the generality of the QCE approach~\eref{eq:scalingNgen}, the shifting procedure (\ref{eq:EshiftChi}--\ref{eq:xequ}), and the subsequent asymptotic analysis, universality \eref{eq:universalchigen} of $\chi$ is not restricted to contact interaction \eref{eq:Hdelta}.

We close with a few remarks:
(i) Our method provides predictions for regions of excitation spectra and particle numbers that are barely accessible via full numerical calculations.
(ii) Universality \eref{eq:universalchigen} of $\chi$ directly implies, through \eref{eq:NequN}, universal features for ${\cal N}_\alpha(E) \!=\! {\cal N}_0(E-\chi\Delta_\infty)$ and for the MB DOS $\rho^{(N)}(E) \! =\! \rho_0^{(N)}(E\!-\!\chi\Delta_\infty) [1- \rmd(\chi \Delta_\infty)/\rmd E ]$, both represented in terms of their noninteracting limits at shifted energy.
(iii) Corresponding expressions for the microcanonical and canonical partition functions and thereby thermodynamic quantities follow right away.
E.g., \eref{eq:universalchigen} implies that the microcanonical temperature $T$ can be determined as well by the scaled variables~\eref{eq:scaledvars}. 
Thus, in the thermodynamic limit $N,V_\eff \to \infty$ with $n_\eff$ fixed, partial fermionization $\chi(T,\alpha,n_\eff)$ is an intensive quantity.
(iv) Eq.~\eref{eq:scalingNgen} holds also true for fermions indicating that our approach can be genereralized to fermions.
(v) Another application concerns MB scattering through interacting media due to a fundamental relation between the smooth DOS and the average dwell time \cite{Lyuboshitz1977,Iannaccone1995} that is, in the single-particle case, robust against disorder implying universality \cite{Pierrat2014}.


To conclude we have shown that the consistent use of short-time dynamical information in the description of short-range-interacting 1D systems enables a separation of interaction and confinement effects implying universal features of smoothed MB spectra and related thermodynamic properties. 
The way universality is derived does not depend on details of the short-range interactions and is not restricted to 1D systems.
Hence we envisage a generalization to higher dimensions and other types of interaction.

We acknowledge illuminating discussions with B. Geiger, P. Schmelcher and S. Tomsovic, and partial financial support from the Deutsche Forschungsgemeinschaft through Research Unit FOR760 and project Ri689/14--1.


\renewcommand{\thefigure}{\Alph{section}\arabic{figure}}
\renewcommand{\thesection}{\Alph{section}}
\renewcommand{\thesubsection}{\arabic{subsection}}
\renewcommand{\theequation}{\Alph{section}\arabic{equation}}

\newcommand{\sectionQ}[1]{%
	\stepcounter{section}\setcounter{subsection}{0}\setcounter{figure}{0}\setcounter{equation}{0}%
	\section{\NoCaseChange{\normalsize{Appendix {\thesection}: #1}}}%
}
\newcommand{\subsectionQ}[1]{%
	\stepcounter{subsection}%
	\subsection{{\thesubsection}. #1}%
}

\setcounter{section}{0}

\onecolumngrid

\begin{center}
	\vspace*{8ex}
	\bf \Large Appendix
	\vspace*{3ex}
\end{center}

\twocolumngrid

\sectionQ{Noninteracting case}
\label{sec:sm:nonint}
The QCE for \textit{noninteracting} bosonic or fermionic systems has been addressed in Ref.~\cite{Hummel2014}.
The results needed for the presented work are here briefly recapped for convenience and in order to adapt the notation.
The general form
\begin{equation} \label{eq:sm:ZnonintgenA}
	Z_{0,\pm}^{(N)}(\beta) = \frac{1}{N!} \sum_{\mathfrak{N} \vdash N} (\pm 1)^{N-|\mathfrak{N}|} c^{(N)}_\mathfrak{N} \mathcal{A}_\mathfrak{N}(- \rmi \hbar \beta)
\end{equation}
of the multi-particle partition function fro noninteracting bosons $(+)$ or fermions $(-)$ is based on the cycle decomposition of permutations $P$ involved in the symmetrization of the Hilbert space due to indistinguishability.
A specific clustering, \ie, decomposition of $P$ into cycles of particular lengths $n_i$, is characeterized by the multiset $\mathfrak{N} \!=\!  \{ n_1 , n_2 , \ldots , n_{|\mathfrak{N}|}\}$, summing up to a total of $\sum_{n \in \mathfrak{N}} n = \sum_{i=1}^{|\mathfrak{N}|} n_i = N$.
The sum in Eq.~\eref{eq:sm:ZnonintgenA} thus runs over all partitions of $N$, denoted by $\mathfrak{N} \vdash N$, while $|\mathfrak{N}|$ denotes the number of parts and the combinatorial factor
\begin{equation} \label{eq:sm:c}
	c^{(N)}_\mathfrak{N} := \frac{N!}{\prod_{n\in \mathfrak{N}} n \prod_{n^\prime} m_{\mathfrak{N}}(n^\prime)!}
\end{equation}
is the number of distinct permutations of $N$ with a cycle-decomposition corresponding to $\mathfrak{N}$, where $m_{\mathfrak{N}}(n)$ is the multiplicity of $n$ in $\mathfrak{N}$.

The amplitude of each clustering $\mathfrak{N}$ in Eq.~\eref{eq:sm:ZnonintgenA} is the product
\begin{equation}
	\mathcal{A}_\mathfrak{N}(-\rmi \hbar \beta) = \prod_{n \in \mathfrak{N}} \mathcal{A}_{n}(- \rmi \hbar \beta)
\end{equation}
of the amplitudes $\mathcal{A}_n$ of the individual clusters, specified by their cluster sizes $n$, \ie, the number of particles involved in the cycles.

Using the semigroup property of the single-particle propagator $K^{(1)}_0$ and identifying $\bq_{n+1}\equiv \bq_1\!=\! \bq$ yields the $n$-body cluster amplitude
\begin{equation}
	\mathcal{A}_{n}(t) = \int \rmd^D\!q \, K^{(1)}_0(\bq, \bq;n t) = \mathcal{A}_1(nt) \,.
\end{equation}
Equivalently, the noninteracting $N$ particle partition function \eref{eq:sm:ZnonintgenA} can be written as
\begin{equation} \label{eq:sm:ZnonintgenZ}
	Z_{0,\pm}^{(N)}(\beta) = \frac{1}{N!} \sum_{\mathfrak{N} \vdash N} (\pm 1)^{N-|\mathfrak{N}|} c^{(N)}_\mathfrak{N} \prod_{n \in \mathfrak{N}} Z^{(1)}(n\beta)
\end{equation}
in terms of the single-particle partition function $Z^{(1)}(\beta)$.
For smooth and homogeneously scaling external potentials $\Vext(\bq)$ the latter is evaluated in short-time approximation by replacing
\begin{equation}
	K_0^{(1)}(\bq,\bq;t) \simeq \exp \!\left[-\frac{\rmi}{\hbar}\Vext(\bq)t\right] K_{\rm free}^{(1)}(\bq,\bq;t) \,,
\end{equation}
where $K_{\rm free}$ stands for unconfined propagation; resulting in
\begin{equation} \label{eq:sm:Z1}
	Z^{(1)}(\beta) = \mathcal{A}_1(-\rmi \hbar \beta) = \frac{V_\eff}{\lamT^d} \propto \beta^{-d/2}
\end{equation}
and consequently
\begin{equation} \label{eq:sm:Zn}
	Z^{(1)}(n \beta) = \mathcal{A}_n(-\rmi \hbar \beta) = \frac{V_\eff}{\lamT^d} n^{-d/2} \,.
\end{equation}
The scaling with the effective dimension
\begin{equation} \label{eq:sm:deff}
	d = D+\frac{2}{\mu}D
\end{equation}
with $\mu$ the degree of homogeneity in the external potential $\Vext(\bq) = w^\mu \Vext(\bq / w)$ allows to absorb the dependence on cluster sizes $n \in \mathfrak{N}$ into combinatorial coefficients separated from the dependence on the relevant physical quantities, which are the temperature encoded in the thermal de Broglie wavelength
\begin{equation} \label{eq:sm:lamT}
	\lambda_{\rm T} = \left( \frac{m}{2 \pi \hbar^2 \beta} \right)^{-\frac{1}{2}}
\end{equation}
and the (effective) volume
\begin{equation} \label{eq:sm:Veff}
	V_\eff = \left( \frac{2 \pi \hbar^2}{m e_0} \right)^{D / \mu} \int \rmd^D\!q \, \exp[{-\Vext(\bq)/e_0}]
\end{equation}
with an arbitrary unit of energy $e_0$, which, for $\hbar^2/(2m)=1$, coincides with the definition given in the main text.

The final explicit expression [see \Leref{eq:Znonint}]
\begin{equation} \label{eq:sm:Znonint}
 Z^{(N)}_{0,\pm}(\beta) = \sum_{l=1}^N z^{(N,d)}_{\pm,l} \left( \frac{V_\eff}{\lambda_T^d} \right)^l
\end{equation}
involves the coefficients
\begin{equation} \label{eq:sm:zcoeffunconf}
	z^{(N,d)}_{\pm,l} = (\pm 1)^{N-l} C_l^{(N,d)} / l! \,,
\end{equation}
where the index $l$ corresponds to the number of clusters [$|\mathfrak{N}|$ in Eq.~\eref{eq:sm:ZnonintgenA}] the total number of particles $N$ is divided into, or equivalently the number of cycles in a permutation.
The universal coefficients $C_l^{(N,d)}$ are given by
\begin{equation} \label{eq:sm:univcoeffunconf}
	\begin{split}
	C_l^{(N,d)}
		&{}= \sum_{\substack{\mathfrak{N} \vdash N \\ |\mathfrak{N}| = l}} \frac{l!}{\prod_n m_{\mathfrak{N}}(n)!} \Bigl( \prod_{n\in \mathfrak{N}} \frac{1}{n} \Bigr)^{d/2+1} \\
	&{}= \sum_{{\substack{n_1,\ldots,n_l =1\\ \sum n_k = N}}}^N  \Bigl( \prod_{k=1}^l \frac{1}{n_k} \Bigr)^{d/2+1}
	\end{split}
\end{equation}
and result from the scaling~\eref{eq:sm:Z1} of single-particle partition functions with the effective dimension $d$ together with summing up all contributions with the same number $l$ of clusters, irrespective of their individual sizes, absorbing the combinatorial factors $c^{(N)}_\mathfrak{N}$, Eq.~\eref{eq:sm:c}.
A full combinatorial derivation of Eq.~\eref{eq:sm:univcoeffunconf} as well as a recursive method for fast evalation at larger values of $N$ was given in~\cite{Hummel2014}.

\sectionQ{QCE in thermal equilibrium}%
\label{sec:sm:kernels}%
\subsectionQ{Arbitrary order}%
In full generality, \ie, arbitrary order, dimensionality $D$, and interaction with short-range character, the QCE partition function [see \Leref{eq:Z1}] is entirely determined by interaction kernels $a_n^{\mathfrak{C}}(\beta \alpha)$, each associated with an irreducible cluster of size $n$, uniquely labeled by a symbol $\mathfrak{C}$ and represented by an irreducible diagram [see, \eg, Figs. \ref{fig:diagrams}(b), \ref{fig:diagrams}(e), and \ref{fig:diagrams}(f) of the main text].
The kernels are defined via their unique relation to the amplitudes
\begin{equation} \label{eq:sm:ACscaling}
	\mathcal{A}_n^{\mathfrak{C}}(-\rmi \hbar \beta) = \frac{V_\eff}{\lamT^d} n^{-d/2} a_n^{\mathfrak{C}}(\beta \alpha)
\end{equation}
that denote the value of the corresponding diagrams, where $\alpha$ denotes the energy associated with the coupling strength.
The generic scaling property \eref{eq:sm:ACscaling} can be shown by consistent use of short-time dynamical information \cite{Hummel2017thesis,HummelLong}.
It expresses the fact that interactions only affect the local internal dynamics of a cluster, independent of the external potential $U(\bq)$, while confinement effects are separated, affecting the cluster as a whole.
The information about $U(\bq)$ is carried by the effective volume~\eref{eq:sm:Veff} and the effective dimension~\eref{eq:sm:deff}.
In the general expression for the QCE partition function [\Leref{eq:Z1}] the generic scaling \eref{eq:sm:ACscaling} allows to subsume the effect of the interaction-kernels in the interaction-related terms $ \Delta z_{\pm,l}^{(N,d)}(\beta \alpha) $.
They add functional dependence on $\beta \alpha$ to the coefficients of the corresponding noninteracting partition function [\Leref{eq:Znonint}], while the polynomial structure in $V_\eff \beta^{-d/2}$ is unchanged.
Due to the significance of this scaling we introduce the thermal interaction strength
\begin{equation} \label{eq:sm:s}
	s = \beta \alpha \,.
\end{equation}

\begin{figure}
	\includegraphics[width=\columnwidth]{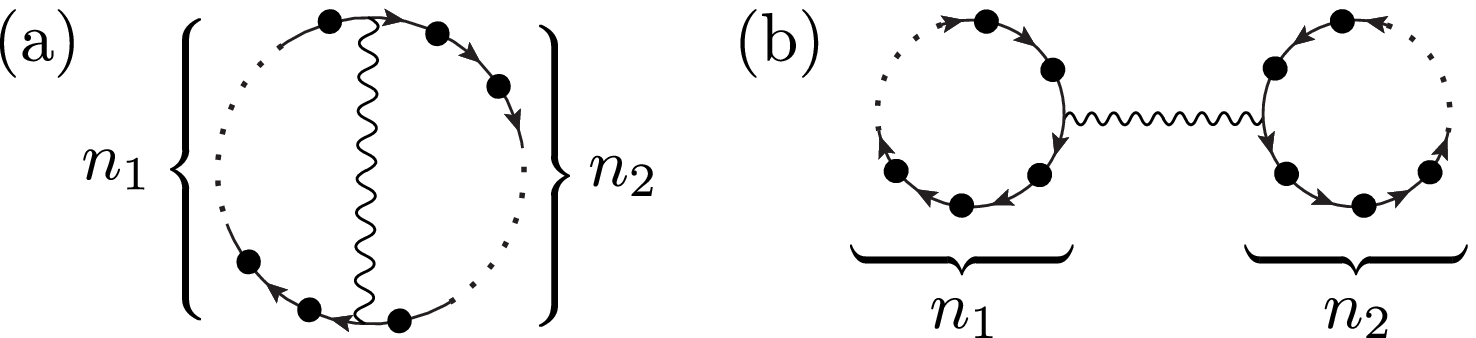}
	\caption{\label{fig:sm:intrainter}
		The two classes of interacting cluster diagrams in first-order QCE:
		(a) The intra-cycle cluster where the interaction effect happens between two particles within the same cycle.
		(b) The inter-cycle structure where the interaction effect between two particles links two distinct cycles together to a single large cluster.
	}
\end{figure}

\subsectionQ{First order}
In first-order QCE, the cluster structure $\mathfrak{C}$ is fully determined by specifying intra- or inter-cycle configuration, the cluster size $n$, and its partition into $n_1$ and $n-n_1$, referring to the location of the interacting pair within the cyclic structure of the permutation (see \fref{fig:sm:intrainter}).
After symmetrization of the inter- and intra-cycle configuration Eq.~\eref{eq:sm:ACscaling} reduces to \Leref{eq:Ascaling}, involving only first-order kernels $ a^{\suppm}_{\subnnm}(s) $.
A combinatorial refinement of \Leref{eq:DeltaZ} yields the relation
\begin{equation} \label{eq:sm:Deltaz}
	\Delta z_{\pm,l}^{(N,d)}(s) = \sum_{n=2}^{N-l+1} \frac{(\pm 1)^n}{n^{d/2}} z_{\pm,l-1}^{(N-n,d)} \sum_{n_1=1}^{n-1} a^{\suppm}_{\subnnm}(s) \,.
\end{equation}

\subsectionQ{Delta-type contact interaction}
We turn now to the case of contact interaction in 1D [see \Leref{eq:Hdelta}].
For two distinguishable particles of equal mass $m$ that live on an infinite line and are interacting via a repulsive Dirac delta pseudo-potential a separation into relative and center-of-mass coordinates allows to relate the interacting two-body propagator
\begin{equation}
	K^{(2)}_0(\bqf, \bqi; t) + \Delta K^{(2)}_\alpha(\bqf, \bqi; t)
\end{equation}
to the known 1D propagator for a single particle on a line with a Dirac delta barrier (see, \eg, \cite{manoukian1989}).
In imaginary time $t = - \rmi \hbar \beta$ and coordinates
\begin{equation}
	\begin{split}
		X^{\mathrm{i},\mathrm{f}} &{}\coloneqq \frac{1}{2} \left( x_1^{\mathrm{i},\mathrm{f}} + x_2^{\mathrm{i},\mathrm{f}} \right) \,, \\
		\Delta x^{\mathrm{i},\mathrm{f}} &{}\coloneqq x_1^{\mathrm{i},\mathrm{f}} - x_2^{\mathrm{i},\mathrm{f}} \,, \\
		x_j &{}\coloneqq \lamT^{-1} q_j
	\end{split}
\end{equation}
that are scaled with the thermal de Broglie wavelength $\lamT$, the interacting part of the two-body propagator reads
\begin{equation} \label{eq:sm:K2deltascaled}
	\begin{split}
		&\Delta K^{(2)}_{\alpha}(\bqf, \bqi; - \rmi \hbar \beta) =
			- \sqrt{s} \lamT^{-2} \rme^{-2\pi (X^{\mathrm{f}}-X^{\mathrm{i}})^2 } \\
		&\quad \times \int_0^\infty \rmd u\;
			\exp\!\left[ -\sqrt{\pi s} u - \frac{\pi}{2} \left( |\Delta x^{\mathrm{f}}|  + |\Delta x^{\mathrm{i}}| + u \right)^2 \right] \,.
	\end{split}
\end{equation}
Using this expression, the value of intra-cluster diagrams $A_{\subnn}^{\rm intra}$ [see~\fref{fig:sm:intrainter}(a), and~\fref{fig:diagrams}(e) of the main text] is found to be
\begin{equation} \label{eq:sm:Aintradelta}
	\begin{split}
		&A_{\subnn}^{\rm intra} = - \frac{L}{\lambda_T n^{1/2}} \frac{\sqrt{2 s}}{4 \pi } 
			\int_0^\infty \rmd r \; \int_{-\infty}^\infty \rmd z \int_0^\infty \rmd u \\
		&\qquad \times \exp \left[-\frac{1}{8} z^2 - \sqrt{\frac{s}{2}} u - \frac{1}{8}( | \nb z + r | + |r| + u )^2 \right] \,,
	\end{split}
\end{equation}
where $L = V_\eff$ is the available length of the 1D system and $n=n_1+n_2$ and $\nb=\sqrt{(2 n_1 n_2)/n - 1}$.

\begin{figure}
	\includegraphics[width=0.8\columnwidth]{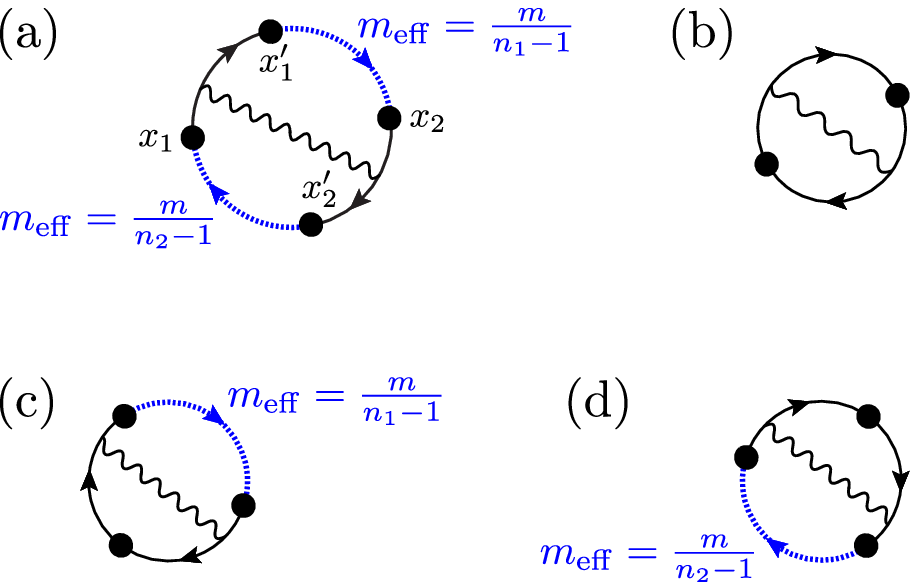}
	\caption{\label{fig:sm:QCE1reducedintra}
		The four cases of reduced effective intra-cluster diagrams (see~\fref{fig:sm:intrainter}a) after convolution of consecutive single-particle propagators.
		(a) $n_{1,2} \geq 2$ reduces to an effective cluster of four constituents with scaled coordinates $x_i, x_i^{\prime}$.
		(b) $n_{1,2} = 1$ is not reduced and remains a two-body cluster.
		(c) $n_1 \geq 2, n_2 = 1$ reduces to an effective three-body cluster as well as the case
		(d) $n_1=1, n_2 \geq 2$.
		Single particle propagators of effective masses are marked with broken blue thick lines.
		All four cases are subsumed by the integral in Eq.~\eref{eq:sm:Aintradelta}.
	}
\end{figure}

Expression~\eref{eq:sm:Aintradelta} is found after using the semigroup convolution property on all consecutive single-particle propagators reducing the cluster diagram to a maximum number of four constituents (see~\fref{fig:sm:QCE1reducedintra}), some of which have altered (effective) masses (or equivalently modified propagation times).
The remaining integration variables refer to the relative coordinate of the two interacting particles scaled with $\lamT$, with the initial distance $r=x_2 - x_1$ and the average distance $z = ((x_2 - x_1) + (x_2^\prime - x_1^\prime)) / \nb$ during the process.

Analogue considerations can be made on the inter-cycle cluster diagrams [see \fref{fig:sm:intrainter}(b) and \fref{fig:diagrams}(f) of the main text] and lead to the exact identity
\begin{equation}
	\mathcal{A}_{\subnn}^{\rm intra} = \mathcal{A}_{\subnn}^{\rm inter} \equiv \mathcal{A}_{\subnn} \,,
\end{equation}
which is a special feature of the delta-type interaction and implies the redundancy of delta-interactions characteristic for \textit{spinless fermions}.
This special implication of the Pauli exclusion principle is thereby confirmed within QCE to first order, instead of being imposed explicitly.
In the cluster expansion this happens by rendering the sum in \Leref{eq:DeltaZ} of all (first-order) interacting irreducible diagrams of a specific size $n$ equal to null, since $\mathcal{A}^{\supm}_{\subnnm} = (\mathcal{A}^{\mathrm{intra}}_{\subnnm} - \mathcal{A}^{\mathrm{inter}}_{\subnnm})/2 = 0$.
While, as a physical fact, this circumstance is expected we find it here as a nontrivial cancellation effect confirming the whole approach.

The bosonic cluster contribution $\mathcal{A}^{\supp} = \mathcal{A} = \mathcal{A}^{\mathrm{intra}}$, Eq.~\eref{eq:sm:Aintradelta}, fulfills the generic scaling \eref{eq:sm:ACscaling} [see also \Leref{eq:Ascaling}], in this case reading
\begin{equation}
	\mathcal{A}^{\supp}_{\subnn}(\alpha, \beta) =  \frac{L}{\lamT}  n^{-1/2} a^{\supp}_{\subnn}(\beta\alpha) \,,
\end{equation}
where further evaluation of Eq.~\eref{eq:sm:Aintradelta} results in the interaction kernel
\begin{equation} \label{eq:sm:adelta2}
	\begin{split}
		\hspace*{-1ex}a_{\subnn}^{\supp}(s) =  &- \frac{2}{\pi} \atan \frac{1}{\nb} + \frac{2}{\sqrt{\pi}} \left[ \frac{\nb^2}{\sqrt{1+\nb^2}} \sqrt{s} \right.\\
		&{}- \nb \sqrt{s} \rme^s \erfc (\sqrt{s}) + \left. (1 - 2 \nb^2 s) F_{\nb}(s)
			\vphantom{\frac{\nb^2}{\sqrt{1+\nb^2}}}
			\right] 
	\end{split}
\end{equation}
with
\begin{equation} \label{eq:sm:F}
	F_{\nb}(s) = \int_0^\infty \rmd z\, \exp \!\left[{-z\left(z + 2\sqrt{(1+\nb^2)s}\right)}\right] \erfc(\nb z) \,.
\end{equation}

Setting $\nb = 0$ one recovers the case involving only two particles $\mathcal{A}_{\subuu} = \frac{L}{\lamT \sqrt{2}} (-1 + \rme^s \erfc (\sqrt{s}))$ on a line \cite{Geiger2017}, also related to a corresponding expression in fully balanced spin-one-half Fermi gases which has been derived in the context of second-order virial expansion \cite{Hoffman2015}.

\sectionQ{QCE in spectral representation}
\label{sec:sm:LCF}
The general relation between the level counting function and the canonical partition function via inverse Laplace transform,
\begin{equation}
	\mathcal{N}_\alpha(E) = \Linv \left[ \frac{1}{\beta} Z_{\alpha,\pm}^{(N)}(\beta) \right](E) \,,
\end{equation}
applied to \Leref{eq:Z1} gives the general QCE expression [\Leref{eq:scalingNgen}] for $\mathcal{N}_\alpha(E)$ with spectral ``coefficient functions''
\begin{equation} \label{eq:sm:ggeneral}
		g_{\pm,l}^{(N,d)}(\epsilon) = \epsilon^{-\frac{ld}{2}} \Linvs \left[ \Delta z^{(N,d)}_{\pm,l}(s) s^{-\frac{ld}{2}-1} \right](\epsilon) \,,
\end{equation}
defined in terms of the interaction kernels $ a_n^{\mathfrak{C}}(s) $ and the effective dimension $d$, where we used the identity
$ \Linv\,[ f(\beta) ](E) = 1 / a \Linvab\,[ f(\beta) ](E/a) $.

We focus on the explicit computation of the spectral coefficients $g_{+,l}^{(N,d)}(\epsilon)$ in first-order QCE for contact interactions (see previous section) by applying Eq.~\eref{eq:sm:ggeneral} to Eq.~\eref{eq:sm:Deltaz} using the explicit kernels \eref{eq:sm:adelta2}.
For more clarity in calculus we split the internal factors of first-order cluster diagrams, according to
\begin{equation}
	a_{\subnnm}^{\supp}(s) = a_1(s,\nb) + a_2(s,\nb) + a_3(s,\nb) + a_4(s,\nb) \,,
\end{equation}
into their four addends
\begin{equation} \label{eq:sm:a1234}
	\begin{split}
		a_1(s,\nb) &= \frac{2}{\pi} \atan{\nb} - 1 + \frac{2 \nb^2}{\sqrt{\pi (1+\nb^2)}} \sqrt{s} \,,\\
		a_2(s,\nb) &= - \frac{2}{\sqrt{\pi}} \nb \sqrt{s} \rme^s \erfc(\sqrt{s}) \,,\\
		a_3(s,\nb) &= \frac{2}{\sqrt{\pi}} F_{\nb}(s) \,,\\
		a_4(s,\nb) &= - \frac{4}{\sqrt{\pi}} \nb^2 s F_{\nb}(s) = - 2 \nb^2 s a_3(s,\nb) \,,
	\end{split}
\end{equation}
where we have absorbed the dependence on $n_1$ and the cluster size $n$ into $\nb$.
Combining the explicit expressions~\eref{eq:sm:a1234} for the contact-interacting 1D Bose gas with the general first-order QCE formula \eref{eq:sm:Deltaz}] for the coefficients $\Delta z^{(N,d)}_{+,l}(s)$ of the partition function and subsequently plugging it into~\Eref{eq:sm:ggeneral} gives
\begin{equation} \label{eq:sm:gfromb}
	g_{+,l}^{(N,d)}(\epsilon) = \sum_{n=2}^{N-l+1} n^{-\frac{d}{2}} z_{+,l-1}^{(N-n,d)} \sum_{n_1=1}^{n-1} \sum_{j=1}^{4} b_j^{(ld)}(\epsilon,\nb)
\end{equation}
with
\begin{equation}
	b_j^{(ld)}(\epsilon,\nb) = \epsilon^{-\frac{ld}{2}} \Linvs\left[ s^{-\frac{ld}{2}-1} a_j(s,\nb) \right](\epsilon) \,.
\end{equation}
In the following explicit expressions for the four $b_j$ are calculated for the case of integer upper index $l d \in \mathbb{N}$.
This covers all possible contributions one can get for integer effective dimension $d \in \mathbb{N}$, Eq.~\eref{eq:sm:deff}, including the important 1D cases of
\begin{itemize}
	\item[\textit{i)}] vanishing external potential $\Vext = 0$ (Lieb-Liniger), where $\mu = \infty$, $d = D = 1$,
	\item[\textit{ii)}] the harmonically trapped Bose gas $\Vext(q) \propto q^2$, where $\mu = 2$, $d = 2 D = 2$, and
	\item[\textit{iii)}] linear potentials like a linear well $\Vext(q) \propto |q|$, where $\mu = 1$, $d = 3 D = 3$.
\end{itemize}
To ease notation for the computations, the effective dimension is assumed to be $d=1$ whithout loss of generality, since all $d \in \mathbb{N}$ are also covered by renaming $l \mapsto l d \in \mathbb{N}$.

\subsectionQ{Calculation of \texorpdfstring{$b_1^{(l)}(\epsilon,\nb)$}{b1}}
Applying standard rules of inverse Laplace transformation to powers of $s$ gives
\begin{equation} \label{eq:sm:Linva1}
	\begin{split}
		b_1^{(l)}(\epsilon,\nb) = &\left( \frac{2}{\pi} \atan{\nb} -1 \right) \frac{\theta(\epsilon)}{\Gamma\left( \frac{l}{2}+1 \right)} \\
		& {}+ \frac{2 \nb^2}{\sqrt{\pi (1+\nb^2)}} \frac{\theta(\epsilon)}{\Gamma\left(\frac{l}{2}+\frac{1}{2}\right) \sqrt{\epsilon} } \,.
	\end{split}
\end{equation}
 
\subsectionQ{Calculation of \texorpdfstring{$b_2^{(l)}(\epsilon,\nb)$}{b2}}
Following the recursive approach in \cite{Geiger2017} gives
	\begin{multline}
		b_2^{(l)}(\epsilon,\nb) = - \frac{2 \nb}{\sqrt{\pi}}
			\frac{ \left(1 + \frac{1}{\epsilon}\right)^{\frac{l}{2}-\frac{1}{2}} }{ \Gamma\left(\frac{l}{2}+\frac{1}{2}\right) \sqrt{\epsilon} } h_\lambda(\epsilon) \\
		{}+ \frac{2 \nb}{\pi} \sum_{k=1}^{\lfloor \frac{l}{2} \rfloor} \frac{\Gamma\left( \frac{l}{2}-k+\frac{1}{2}\right)}
			{ \Gamma\left( \frac{l}{2}-k+1\right) \Gamma\left( \frac{l}{2} + \frac{1}{2} \right) }
			\left( 1 + \frac{1}{\epsilon}\right)^{k-1} \frac{\theta(\epsilon)}{\epsilon} \,,
	\end{multline}
with the definitions
\begin{equation} \label{eq:sm:h}
	h_\lambda(\epsilon) = \begin{cases}
																\frac{2}{\pi} \theta(\epsilon) \atan ( \sqrt{\epsilon} ) & : \quad \lambda = \frac{1}{2} \,,\\
																\theta(\epsilon) & : \quad \lambda = 0 \,,
	                             \end{cases}
\end{equation}
and
\begin{equation}
	\lambda = \frac{1}{2} ( l\ {\rm mod}\ 2 ) = \begin{cases}
																												\frac{1}{2} & : \quad l\ {\rm odd} \,,\\
																												0 & : \quad l\ {\rm even} \,.\\
	                                                     \end{cases}
\end{equation}
Here $\lfloor q \rfloor$ denotes the integer $n \leq q$ that is closest to $q$.
\\

\subsectionQ{Calculation of \texorpdfstring{$b_3^{(l)}(\epsilon,\nb)$}{b3}}{}
To simplify the following analysis we define
\begin{equation}
	\tilde{F}_{\nb}(s) := \rme^{-(1+\nb^2)s} F_{\nb}(s) \,.
\end{equation}
The integral in $\tilde{F}_{\nb}(s)$ can not be evaluated to elementary expressions directly.
In contrast to that its inverse Laplace transform can be related to the solvable derivative given by
\begin{equation} \label{eq:sm:Fprime}
	\rme^{(1+\nb^2)s} \tilde{F}_{\nb}^\prime(s) = \frac{\nb}{2} s^{-\frac{1}{2}} \rme^s \erfc(\sqrt{s}) - \frac{1}{2} \sqrt{1+\nb^2} s^{-\frac{1}{2}} \,.
\end{equation}
Using this observation we calculate
\begin{align}
	\Linvs\left[F_{\nb}(s)\right](\epsilon) &= \Linvs\left[\tilde{F}_{\nb}(s)\right](\epsilon + (1+\nb^2)) \nonumber \\
	&{}= -\frac{ \Linvs\left[\tilde{F}_{\nb}^\prime(s)\right](\epsilon+(1+\nb^2)) }{ \epsilon + (1+\nb^2) } \nonumber \\
	&{}= -\frac{ \Linvs\left[\rme^{(1+\nb^2)s}\tilde{F}_{\nb}^\prime(s)\right](\epsilon) }{ \epsilon + (1+\nb^2) } \nonumber \\
	&{}= (\epsilon + (1+\nb^2))^{-1} \nonumber \\
	&\quad\times\left( \frac{\sqrt{1+\nb^2}}{2\sqrt{\pi}}  \frac{\theta(\epsilon)}{\sqrt{\epsilon}}
		- \frac{\nb}{2\sqrt{\pi}} \frac{\theta(\epsilon)}{\sqrt{1+\epsilon}} \right) \,.
\end{align}
From there we get
\begin{align} \label{eq:sm:InitInt}
	\Linvs\left[s^{-1}F_{\nb}(s)\right](\epsilon) &= \int_{-\infty}^{\epsilon} \rmd x \, \Linvs\left[F_{\nb}(s)\right](x) \nonumber \\
	&= \frac{\theta(\epsilon)}{\sqrt{\pi}} \left[ \atan\left( \sqrt{\frac{\epsilon}{1+\nb^2}} \right) \right. \nonumber \\
	& \qquad\quad\left. + \atan\left( \sqrt{\frac{\nb^2}{1+\epsilon}} \right) - \atan \nb \right] \,,
\end{align}
and
\begin{equation} \label{eq:sm:InitHalfInt}
	\begin{split}
		\MoveEqLeft \Linvs\left[s^{-\frac{1}{2}}F_{\nb}(s)\right](\epsilon) \\
		&=\int_{-\infty}^\infty \rmd x \, \Linvs\left[s^{-\frac{1}{2}}\right](\epsilon - x) \Linvs\left[F_{\nb}(s)\right](x) \\
		&\begin{split}
			{}=\frac{\theta(\epsilon)}{2 \pi} \int_0^{\epsilon} \rmd x \, \frac{1}{\sqrt{\epsilon-x}} &\left[ \frac{\sqrt{1+\nb^2}}{\sqrt{x}(x+(1+\nb^2))} \right. \\
			&\left. {} - \frac{\nb}{\sqrt{1+x}(x+(1+\nb^2))} \right]
		\end{split} \\
		&{}= \frac{\theta(\epsilon)}{\pi} (\epsilon + (1+\nb^2))^{-\frac{1}{2}} \atan\left( \frac{1}{\nb} \sqrt{1+\frac{1+\nb^2}{\epsilon}} \right) \,.
	\end{split}
\end{equation}
We calculate $\Linvs\left[s^{-n} \tilde{F}_{\nb}(s)\right]$ for larger negative powers of $s$ using a recursive approach, where Eqs.~\eref{eq:sm:InitInt} and~\eref{eq:sm:InitHalfInt} will serve as initial values.
We define
\begin{equation} \label{eq:sm:DefGn}
	G_n(s) := \Gamma(n) s^{-n} \tilde{F}_{\nb}(s) \,,
\end{equation}
where $n$ may be either integer or half-integer.
Taking the derivative of Eq.~\eref{eq:sm:DefGn} with respect to $s$ leads to
\begin{equation}
	G_{n+1}(s) = - \frac{\partial}{\partial s} G_n(s) + \Gamma(n) s^{-n} \tilde{F}_{\nb}^\prime(s) \,,
\end{equation}
which implies the recursion relation
\begin{equation} \label{eq:sm:RecRel}
	\begin{split}
		\Linvs\left[G_{n+1}(s)\right](\epsilon) ={} &\epsilon \Linvs\left[G_n(s)\right](\epsilon) \\
		& {}+ \Gamma(n) \Linvs\left[s^{-n} \tilde{F}_{\nb}^\prime(s)\right](\epsilon)
	\end{split}
\end{equation}
for the inverse Laplace transformed objects, where the initial values $\Linvs\left[G_1(s)\right]$ or $\Linvs\left[G_{\frac{1}{2}}(s)\right]$ are given explicitly by Eqs.~\eref{eq:sm:InitInt} and~\eref{eq:sm:InitHalfInt}.
The solution to Eq.~\eref{eq:sm:RecRel} is either given by
\begin{equation} \label{eq:sm:GsolInt}
	\begin{split}
		\Linvs\left[G_{n+1}(s)\right](\epsilon) ={} &\epsilon^n \Linvs\left[G_1(s)\right](\epsilon) \\
		& {}+ \sum_{k=1}^n \epsilon^{n-k} \Gamma(k) \Linvs\left[s^{-k} \tilde{F}_{\nb}^\prime(s)\right](\epsilon)
	\end{split}
\end{equation}
for integer indexes or by
\begin{align} \label{eq:sm:GsolHalfInt}
		&\Linvs\left[G_{n+\frac{1}{2}}(s)\right](\epsilon) = \epsilon^n \Linvs\left[G_\frac{1}{2}(s)\right](\epsilon) \nonumber\\
		& {}\quad+ \sum_{k=0}^{n-1} \epsilon^{n-1-k} \Gamma\left(k+\frac{1}{2}\right) \Linvs\left[s^{-k-\frac{1}{2}} \tilde{F}_{\nb}^\prime(s)\right](\epsilon)
\end{align}
for half-integer indexes.
In the given form, both solutions~\eref{eq:sm:GsolInt} and~\eref{eq:sm:GsolHalfInt} are valid for $n \in \mathbb{N}_0$.
After reintroducing the exponential prefactor Eqs.~\eref{eq:sm:GsolInt} and~\eref{eq:sm:GsolHalfInt} become
\begin{align}
		&\Gamma(n+1) \Linvs\left[s^{-n-1} F_{\nb}(s)\right](\epsilon) \nonumber\\
		&\hspace*{1em}{}= (\epsilon + (1+\nb^2))^n \Linvs\left[s^{-1} F_{\nb}(s)\right](\epsilon) \nonumber\\
			&\hspace*{1em}\quad + \sum_{k=1}^n (\epsilon+(1+\nb^2))^{n-k} \Gamma(k) \\
			&\hspace*{5em}\times\Linvs\left[s^{-k} \rme^{(1+\nb^2)s} \tilde{F}_{\nb}^\prime(s)\right](\epsilon) \,,\nonumber
\end{align}
and
\begin{equation}
	\begin{split}
		&\Gamma\!\left( n+\frac{1}{2}\right) \Linvs\left[s^{-n-\frac{1}{2}} F_{\nb}(s)\right](\epsilon) \\
		&\qquad\begin{split}
			&{}= \sqrt{\pi} (\epsilon + (1+\nb^2))^n \Linvs\left[s^{-\frac{1}{2}} F_{\nb}(s)\right](\epsilon) \\
			&\quad\begin{split}
				{}+ \sum_{k=1}^n &(\epsilon+(1+\nb^2))^{n-k} \Gamma\!\left(k-\frac{1}{2}\right) \\
				&\times \Linvs\left[s^{-k+\frac{1}{2}} \rme^{(1+\nb^2)s} \tilde{F}_{\nb}^\prime(s)\right](\epsilon) \,,
			\end{split}
		\end{split}
	\end{split}
\end{equation}
where  $n \in \mathbb{N}_0$.
The remaining step is to calculate $\Linvs\left[s^{-n} \rme^{(1+\nb^2)s} \tilde{F}_{\nb}^\prime(s)\right](\epsilon)$ for $n$ being either integer or half-integer.
Using Eq.~\eref{eq:sm:Fprime} leads to
\begin{equation}
	\begin{split}
		\MoveEqLeft[4] \Linvs\left[s^{-n} \rme^{(1+\nb^2)s} \tilde{F}_{\nb}^\prime(s)\right](\epsilon) \\
		{}={} &\frac{\nb}{2} \Linvs\left[ s^{-n-1} \sqrt{s} \erfc(\sqrt{s}) \right](\epsilon) \\
		&{}- \frac{1}{2} \sqrt{1+\nb^2} \Linvs\left[ s^{-n-\frac{1}{2}} \right](\epsilon) \\
		{}={} &- \frac{\sqrt{\pi}}{4} \epsilon^n b_2^{(2n)}(\epsilon,\nb)
			- \frac{\sqrt{1+\nb^2}}{2 \Gamma(n+\frac{1}{2})} \epsilon^{n-\frac{1}{2}} \theta(\epsilon) \,.
	\end{split}
\end{equation}

For $l \geq -1$ we get
\begin{widetext}
	\begin{equation}\label{eq:sm:b3}
		\begin{split}
			\MoveEqLeft[6] b_3^{(l)}(\epsilon,\nb) = \frac{ \left(1+\frac{1+\nb^2}{\epsilon}\right)^{\frac{l}{2}} }{ \Gamma\!\left( \frac{l}{2} + 1 \right) }
				\left[ t_\lambda(\epsilon,\nb) - \frac{1}{\sqrt{\pi}} \sum_{k=1}^{\lceil \frac{l}{2} \rceil}
				\Gamma(k-\lambda) \left(1+\frac{1+\nb^2}{\epsilon}\right)^{\lambda-k} \left( \frac{\sqrt{\pi}}{2} b_2^{(2(k-\lambda))}(\epsilon,\nb) + \frac{\sqrt{1+\nb^2}}{\Gamma\!\left(k-\lambda+\frac{1}{2}\right)} \frac{\theta(\epsilon)}{\sqrt{\epsilon}} \right) \right] \,,
		\end{split}
	\end{equation}
\end{widetext}
where $\lceil q \rceil$ denotes the integer $n \geq q$ that is closest to $q$ and the function $t_\lambda$ is defined as
\begin{equation}
	t_\lambda(\epsilon,\nb) = \begin{cases}
																\frac{2}{\pi} \theta(\epsilon) \atan\left( \frac{1}{\nb} \sqrt{1+\frac{1+\nb^2}{\epsilon}} \right) & : \lambda = \frac{1}{2} \,, \\
																\frac{2}{\pi} \theta(\epsilon) \left[ \atan\left( \sqrt{\frac{\epsilon}{1+\nb^2}} \right) \right. & \\
																	\qquad \quad \left. {}+ \atan\left( \sqrt{\frac{\nb^2}{1+\epsilon}} \right) - \atan \nb \right] & : \lambda = 0 \,.\\
	                             \end{cases}
\end{equation}

\subsectionQ{Calculation of \texorpdfstring{$b_4^{(l)}(\epsilon,\nb)$}{b4}}

Since Eq.~\eref{eq:sm:b3} is not only valid for $l \in \mathbb{N}$ but also for the values $l=-1,0$ we can use the simple relation between $a_3$ and $a_4$ [see Eq.~\eref{eq:sm:a1234}] to get
\begin{equation}
	b_4^{(l)}(\epsilon,\nb) = - \frac{2 \nb^2 }{\epsilon} b_3^{(l-2)}(\epsilon,\nb)
\end{equation}
for all $l \in \mathbb{N}$.

\newpage
\sectionQ{QCE in fermionization regime---Strong-coupling expansion}
\label{sec:sm:strong}
We employ an exact mapping \cite{cheon1999} of 1D bosonic systems with delta interaction to spinless fermionic systems with an effective attractive zero-range interaction potential, here referred to as \textit{anti delta}.
Application of first-order QCE to the effective fermionic theory relies on the two-body propagator for the anti delta interaction.
To relate it to the propagator in the original system we define, for any two-body propagator $K$, the swapping operation, denoted by $\bar{K}$, as
\begin{equation}
	\begin{split}
		&\bar{K}((q_1',q_2'),(q_1,q_2)) \\
		&\;=\begin{cases}
			K((q_1',q_2'),(q_1,q_2)), &\text{for } (q_1-q_2)(q_1'-q_2')>0 ,\\
			-K((q_1',q_2'),(q_1,q_2)), &\text{for } (q_1-q_2)(q_1'-q_2')<0 \,,
		\end{cases}
	\end{split}
\end{equation}
which gives a relative sign inversion when the two particles have to cross each other along any classical path from $(q_1,q_2)$ to $(q_1',q_2')$.
The interacting propagator $K$ of two distinguishable particles subject to delta interaction is built from its symmetric part $K_+$ and its antisymmetric part $K_-$ \wrt particle exchange,
\begin{equation}
	K = K_+ + K_- \,,
\end{equation}
where $K_+$($K_-$) is defined by all symmetric(antisymmetric) eigenfunctions $\psi_\pm(R,r)$ of the two-body system, where $R,r$ denote center-of-mass and relative coordinates, respectively.
The delta interaction only has an effect on the symmetric wavefunctions $\psi_+(R,r)$, whereas the antisymmetric ones are unaffected $\psi_-(R,r) = \psi_{0,-}(R,r)$, thus we write
\begin{align}
	K_+ &= K_{0,+} + K_\alpha \,,\\
	K_- &= K_{0,-} \,,
\end{align}
where $K_{0,\pm}$ denotes the (anti)symmetric part of the noninteracting propagator and $K_\alpha$ the modification to the symmetric part due to finite interaction.

For the anti delta interaction (which will be denoted by a tilde) the opposite is the case and one has unaffected symmetric wavefunction $\tilde{\psi}_+(R,r) = \psi_{0,+}(R,r)$ whereas the antisymmetric wavefunctions $\tilde{\psi}_-(R,r)$ feel the interaction in form of a jump discontinuity at vanishing relative distance $r$ of the particles.
Because of the exact mapping, those antisymmetric wavefunctions are equivalent with the symmetric ones for the delta interaction with a conditional sign inversion
\begin{equation}
	\tilde{\psi}_-(R,r) = \mathrm{sign}(r) \psi_+(R,r) \,,
\end{equation}
reflected in the propagator $\tilde{K}$ of two distinguishable particles with anti delta interaction as
\begin{equation}
	\begin{split}
		\tilde{K} &= K_{0,+} + \bar{K}_+ \\
		&= K_{0,+} + \bar{K}_{0,+} + \bar{K}_\alpha \,.
	\end{split}
\end{equation}
For first-order QCE calculations one needs then only the modification $\tilde{K}_\alpha$ of the porpagator due to anti delta interaction, thus we write
\begin{equation}
	\begin{split}
		\tilde{K} &= K_0 + \tilde{K}_\alpha \\
		&= K_{0,+} + K_{0,-} + \tilde{K}_\alpha \,,
	\end{split}
\end{equation}
and obtain the final result
\begin{equation} \label{eq:sm:Ferm:prop}
	\tilde{K}_\alpha = \bar{K}_{0,+} + \bar{K}_\alpha - K_{0,-} \,.
\end{equation}
A simple test of this result can be done in the limit $\alpha\rightarrow\infty$ where the symmetric propagator for delta interaction becomes just the swapped version of the free antisymmetric propagator
\begin{equation}
	K_{0,+} + K_\alpha \xrightarrow[\alpha\rightarrow\infty]{} \bar{K}_{0,-} \,,
\end{equation}
so that 
\begin{equation}
	\tilde{K}_\alpha \xrightarrow[\alpha\rightarrow\infty]{} 0 \,,
\end{equation}
which means the fermionic theory is noninteracting in this limit, confirming the fermionization effect \cite{Tonks1936,Girardeau1960}.

Using the relation~\eref{eq:sm:Ferm:prop} in the calculation of the corresponding QCE diagrams involved in the cluster contribution $\tilde{A}_{\subnnm}(s)$ for the fermionic theory one gets then a replacement of the functions $a_{\subnnm} \mapsto \tilde{a}_{\subnnm}$ given by [see Eq.~\eref{eq:sm:a1234} for comparison]
\begin{equation} \label{eq:sm:a1234ferm}
	\begin{split}
		\tilde{a}_1(s,\nb) &= -\frac{2}{\pi} \frac{\nb}{1+\nb^2} - \frac{2 \nb^2}{\sqrt{\pi (1+\nb^2)}} \sqrt{s} \,,\\
		\tilde{a}_2(s,\nb) &= \frac{2}{\sqrt{\pi}} \nb \sqrt{s} \rme^s \erfc(\sqrt{s}) = - a_2(s,\nb) \,,\\
		\tilde{a}_3(s,\nb) &= \frac{2}{\sqrt{\pi}} F_{\nb}(s) = a_3(s,\nb)\,,\\
		\tilde{a}_4(s,\nb) &= \frac{4}{\sqrt{\pi}} \nb^2 s F_{\nb}(s) = - a_4(s,\nb) \,,
	\end{split}
\end{equation}
and consequently 
\begin{equation} \label{eq:sm:b1234ferm}
	\begin{split}
		&\begin{split}
			\tilde{b}^{(l)}_1(\epsilon,\nb) ={} &-\frac{2}{\pi} \frac{\nb}{1+\nb^2} \frac{\theta(\epsilon)}{\Gamma(\frac{l}{2}+1)} \\
			&-\frac{2 \nb^2}{\sqrt{\pi (1+\nb^2)}} \frac{\theta(\epsilon)}{\Gamma(\frac{l}{2}+\frac{1}{2}) \sqrt{\epsilon}} \,,
		\end{split}\\
		&\tilde{b}^{(l)}_2(\epsilon,\nb) = -b^{(l)}_2(\epsilon,\nb) \,,\\
		&\tilde{b}^{(l)}_3(\epsilon,\nb) = b^{(l)}_3(\epsilon,\nb)\,,\\
		&\tilde{b}^{(l)}_4(\epsilon,\nb) = -b^{(l)}_4(\epsilon,\nb) \,,
	\end{split}
\end{equation}
which can then be used in the fermionic version
\begin{equation}
	\tilde{g}_{-,l}^{(N,d)}(\epsilon) = \sum_{n=2}^{N-l+1} n^{-\frac{d}{2}} z_{-,l-1}^{(N-n,d)} \sum_{n_1=1}^{n-1} \sum_{j=1}^{4} \tilde{b}_j^{(ld)}(\epsilon,\nb)
\end{equation}
of the coefficients~\eref{eq:sm:gfromb} modified to anti delta interaction, plugged into the fermionic counting function [\Leref{eq:scalingNgen}] together with the noninteracting fermionic coefficients \eref{eq:sm:zcoeffunconf}
\begin{equation}
	z_{-,l}^{(n,d)} = (-1)^{n-l} z_{+,l}^{(n,d)}
\end{equation}
to get the corresponding counting functions for the fermionization regime.

\begin{figure}
	\includegraphics[width=0.6\columnwidth]{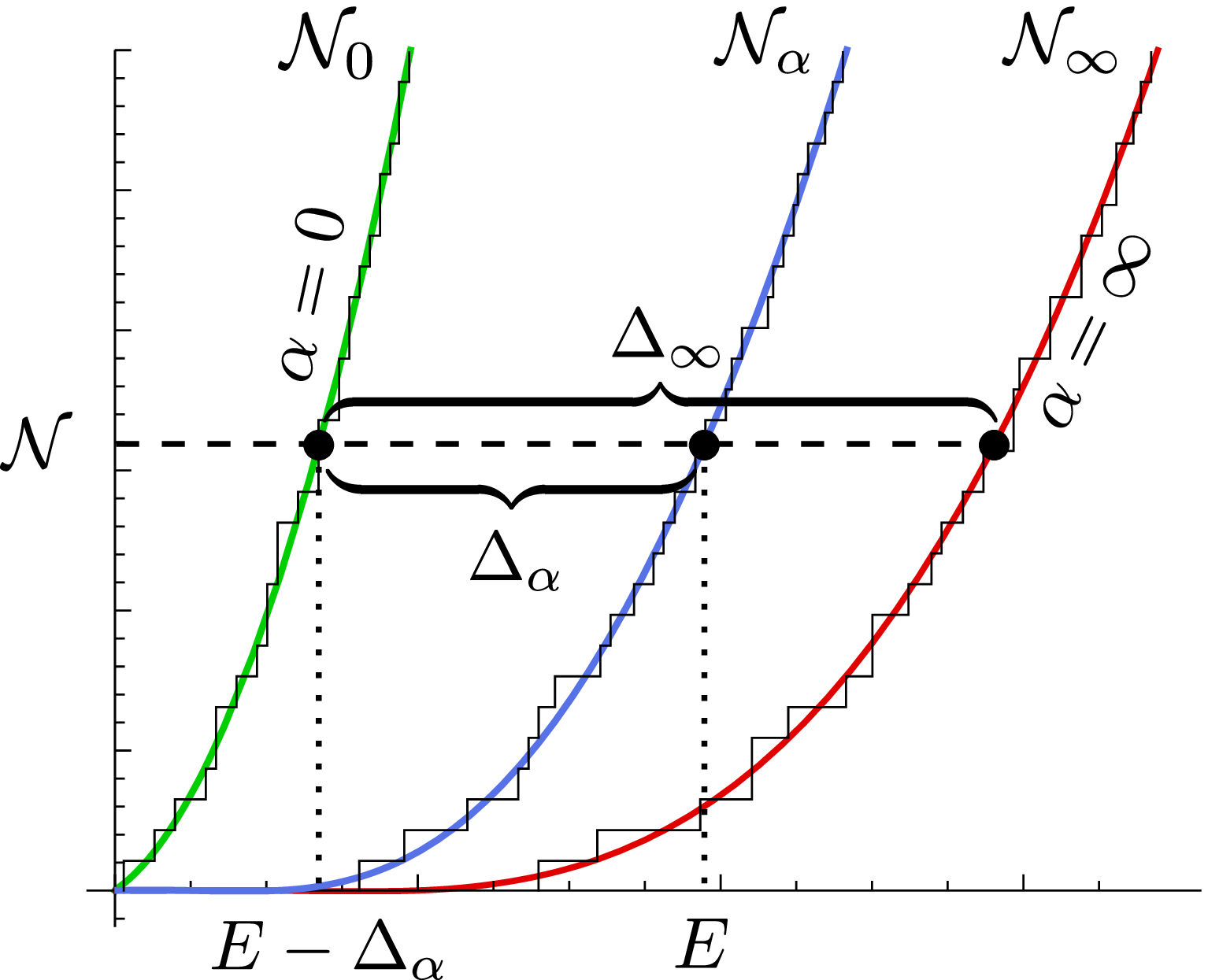}
	\caption{\label{fig:sm:eshiftscheme}
		Schematic method of energy shifting.
	}
\end{figure}

\sectionQ{Energy shifting method}
\label{sec:sm:shifting1}
We write the ansatz of the shifting method, \ie, expressing the effect of interactions as a shift in all energies (see~\fref{fig:sm:eshiftscheme}), as
\begin{equation} \label{eq:sm:Eshiftgeneral}
	\mathcal{N}_\alpha(E) = \mathcal{N}_0(E-\Delta_\alpha) \,.
\end{equation}
The energy shifts $\Delta_\alpha$ are in general allowed to vary with their location in the spectrum.
Besides $E$, also $d, V_\eff, N$, and $\alpha$ are free variables in the formal definition \eref{eq:sm:Eshiftgeneral} of $\Delta_\alpha$, which can be thought of as a reformulation of the problem to find $\mathcal{N}_\alpha(E)$:
Provided that the noninteracting smooth spectrum (represented by $\mathcal{N}_0(E)$) is known, $\Delta_\alpha$ contains all information to construct the interacting one (represented by $\mathcal{N}_\alpha(E)$).
As is pointed out in the following, the shifts $\Delta_\alpha$ have a systematic form induced by the QCE for $\mathcal{N}_0$ and $\mathcal{N}_\alpha$ which allows then for an iterative approximation in a controlled way.
This way one obtains, via Eq.~\eref{eq:sm:Eshiftgeneral}, a method to determine $\mathcal{N}_\alpha$ in an approximative order-by-order manner that is far more efficient than the direct approximation of $\mathcal{N}_\alpha$ using QCE up to a specific order.

\subsectionQ{The full shifts---Infinite coupling}
For the purpose of this section we write the noninteracting counting function as
\begin{equation} \label{eq:sm:LCFnonint}
	\mathcal{N}_0(\tilde{E}) = c_N \tilde{E}^{N d/2} + c_{N-1} \tilde{E}^{(N-1)d/2} + \ldots \,,
\end{equation}
with the dimensionless total energy
\begin{equation}
	\tilde{E} = \rhoeff E \,,
\end{equation}
measured in units of the characteristic energy
\begin{equation}\label{eq:sm:rhoeff}
	\rhoeff^{-1} = \frac{2\pi\hbar^2}{m} V_\eff^{-2/d}
\end{equation}
related to the (effective) system size $V_\eff$ \eref{eq:sm:Veff} and using the short-hand notation
\begin{equation} \label{eq:sm:clShifting}
	c_l = \frac{z_{+,l}^{(N,d)}}{\Gamma\!\left( \frac{l d}{2} + 1\right)}
\end{equation}
for the constant coefficients.
We note that, while we explicitly focus on bosonic systems here, the approach can be easily applied to fermionic systems, replacing $z_+ \mapsto z_-$ in Eq.~\eref{eq:sm:clShifting}, as well.

To start with, we consider 1D bosonic systems with contact interactions and in particular focus on the fermionization limit $\alpha \to \infty$, which can be mapped to noninteracting spinless fermions \cite{Tonks1936,Girardeau1960}.
We thus demand equivalence with the noninteracting fermionic counting function
\begin{equation} \label{eq:sm:ESNinfty}
	\mathcal{N}_\infty( E ) \stackrel{!}{=} c_N \tilde{E}^{N d/2} - c_{N-1} \tilde{E}^{(N-1)d/2} + \ldots \,,
\end{equation}
invoking a requirement on the full shift $\Delta_\infty = \lim_{\alpha\to \infty} \Delta_\alpha$ by comparison \eref{eq:sm:Eshiftgeneral} with the shifted noninteracting counting function.
Up to next-to-leading order in the regime of high energy---or equivalently high temperature and weak quantum degeneracy---the latter reads
\begin{equation} \label{eq:sm:ESN0shift}
	\begin{split}
		&\mathcal{N}_0( E - \Delta_\infty) \\
		&{}= c_N \tilde{E}^{N d/2} \left( 1 - \tilde{E}^{-1} \tilde{\Delta}_\infty \right)^{Nd/2} + c_{N-1} \tilde{E}^{(N-1)d /2} + \ldots \\
		&\begin{split}
			{}= c_N \tilde{E}^{N d/2} &{}- \frac{Nd}{2} c_N \left( \tilde{\Delta}_\infty \tilde{E}^{N d/2 - 1} \right) \\
			&{}+ c_{N-1} \tilde{E}^{(N-1)d /2} + \ldots
		\end{split}
	\end{split}
\end{equation}
with the dimensionless full energy shift defined as
\begin{equation}
	\tilde{\Delta}_\infty = \rhoeff \Delta_\infty \,.
\end{equation}

Equating Eq.~\eref{eq:sm:ESN0shift} with Eq.~\eref{eq:sm:ESNinfty} the second term in the expansion~\eref{eq:sm:ESN0shift} has to contribute to the term of order $\mathcal{O}(\tilde{E}^{(N-1)d/2})$ to correct its sign.
Here we find the case $d=2$ to be of special simplicity, because it corresponds to a full shift $\Delta_\infty$ that is a constant.
From comparison one easily finds in this case
\begin{equation} \label{eq:sm:ESNinftyd2}
	\tilde{\Delta}_\infty = \frac{2 c_{N-1}}{N c_{N}} =  \frac{1}{2}N (N-1) \quad \text{for} \quad d=2\,,
\end{equation}
which is remarkably accurate when applied to the 1D harmonic trapping (for which $d=2$), where the exact spectra of noninteracting bosons and fermions are related level-by-level by an exact constant shift coinciding with the prediction~\eref{eq:sm:ESNinftyd2} from the universal consideration of smooth spectra.

For arbitrary effective dimensionality $d$ the matching~\eref{eq:sm:ESNinfty} demands that the full shifts asymptotically behave like
\begin{equation} \label{eq:sm:ESfullshiftE}
	\tilde{\Delta}_\infty \sim \mathrm{const.}\times\tilde{E}^{1-d/2} \quad \mathrm{as} \quad \tilde{E} \to \infty \,.
\end{equation}
By comparing ideal bosonic and fermionic smooth counting functions for several different values of the effective dimension $d$ and numbers of particles $N$ one finds that the prescription
\begin{equation} \label{eq:sm:ESfullshiftN}
	\tilde{\Delta}_\infty = \mathrm{const.} \times [\mathcal{N}]^{(2/d-1)/N}
\end{equation}
is way more favorable than an energy-dependent prescription as in Eq.~\eref{eq:sm:ESfullshiftE}.
It effectively (in a smooth way) expresses the full fermionization shift of each level as a function of its quantum number $\mathcal{N}$ rather than its (fermionized) energy $E$, related here by $\mathcal{N} = \mathcal{N}_\infty(E)$.
While exhibiting the same correct asymptotics for large $\tilde{E}$ the $\mathcal{N}$-dependent prescription~\eref{eq:sm:ESfullshiftN} quite accurately produces the fully fermionized limit on all energy scales.
So far the explicit $\mathcal{N}$-prescription, Eq.~\eref{eq:sm:ESfullshiftN}, is heuristic (if $d\neq 2$) and only motivated by observation.
For finite coupling $\alpha$ on the other hand, the \textit{partial} shift $\Delta_\alpha$ also involves the infinite shift $\Delta_\infty$ (see subsequent subsection).
There, expressing $\Delta_\infty$ as a function of $\mathcal{N} = \mathcal{N}_\alpha(E)$ is crucial and can be justified by consistency \wrt ``interaction flow'' (see the subsection after the next).

A direct comparison of the next-to-leading order terms in Eqs.~\eref{eq:sm:ESN0shift} and~\eref{eq:sm:ESNinfty}, using $\mathcal{N} = \mathcal{N}_\infty(E)$, yields the explicit full shift
\begin{equation} \label{eq:sm:DEinftymatched}
	\rhoeff \Delta_\infty = \underbrace{\frac{4}{N d} c_{N-1} c_{N}^{-1-(2/d-1)/N}}_{\text{\normalsize const.}} [\mathcal{N}]^{(2/d-1)/N} \,.
\end{equation}

\subsectionQ{The partial shifts---Arbitrary coupling strengths}
The next step in extending the shifts to arbitrary interaction strength crucially relies on the generic structure [\Leref{eq:scalingNgen}] of the QCE:
The effect of arbitrary $\alpha$ is a correction of the coefficients $c_l$ by functions~\eref{eq:sm:ggeneral} of the energy in units of the coupling parameter, $E / \alpha$, while keeping the polynomial structure in $V_\eff E^{d/2}$, or equivalently $\tilde{E}^{d/2}$.
Therefore we apply the same separation in the two distinct energy scales given by $\alpha$ and $V_\eff$ to the shifts.
The corresponding ansatz reads
\begin{equation} \label{eq:sm:ESpartialshift}
	\Delta_\alpha = \chi^{(N,d)}\!\left(\frac{E}{\alpha}\right) \Delta_\infty \,,
\end{equation}
where the partial shift as fraction $\chi \in [0,1]$ of the full shift is a function of the energy in units of the coupling strength, $E/\alpha$, not involving the energy scale given by $V_\eff$.
The system size on the other hand enters the full shift $\Delta_\infty$ (see also \fref{fig:sm:eshiftscheme}), which should in contrast be independent of $\alpha$ in the following sense.
For finite $\alpha$, $\Delta_\infty$ refers to the ``horizontal'' line which should not change while the interacting energy $E$ moves along the horizontal line when changing $\alpha$.
To illuminate this meaning of the separation of energy scales and the prescription~\eref{eq:sm:ESfullshiftN} a bit more, consider the shift of a single exact MB level.
Say, we talk about the $n$-th excited state $E^{(n)}(\alpha)$, start from its noninteracting energy $E^{(n)}(0)$ and adiabatically switch on interactions.
The full shift~\eref{eq:sm:ESfullshiftN} is expressed in terms of the counting function $\mathcal{N} = \mathcal{N}_\alpha(E^{(n)}(\alpha))$ which simply becomes the quantum number $n$.
It does not change during the process of turning on interactions.
The ansatz~\eref{eq:sm:ESpartialshift} then suggests that the ratio $E^{(n)}(\alpha) / \alpha$ of the actual interacting energy of this level and the energy-like coupling parameter is in unique relation to how far this level has moved from the noninteracting $E^{(n)}(0)$ to the fully fermionized counter-part $E^{(n)}(\infty)$.
This notion of invariance of $\Delta_\infty$ during a change of $\alpha$ can be expressed by writing it as a function of the \textit{noninteracting} energy $E_0 \equiv E - \Delta_\alpha$ to eliminate an explicit $\alpha$-dependence.
By using $\mathcal{N} = \mathcal{N}_\alpha(E) = \mathcal{N}_0(E_0)$ in Eq.~\eref{eq:sm:ESfullshiftN} one may write
\begin{equation}
	\Delta_\infty^{(N,d)}(E,\alpha,V_\eff) = \Delta_\infty^{(N,d)}(E_0,V_\eff) \,.
\end{equation}

From the analytical point of view the ansatz~\eref{eq:sm:ESpartialshift} allows a matching in the next-to-leading order term of the expansion in $\tilde{E}$.
The separation into two different energy scales is here crucial.
Matching coefficients in a power expansion in $\tilde{E}$ can be thought of as matching power expansions in the (effective) volume $V_\eff$, while $E/\alpha$ is considered as an independent variable.
This way of separating combines the high amount of analytical control one has over power series expansions with the high value of a nonperturbative description in the interaction.

We briefly comment on the generality of the approach.
The interpretation of $\Delta_\infty$, Eq.~\eref{eq:sm:DEinftymatched}, as full shifts $\lim_{\alpha \to \infty} \Delta_\alpha$ between noninteracting levels and infinite strong coupling or \textit{fermionized} energies only applies to the 1D bosonic case with contact interactions. 
Nevertheless, for arbitrary particle exchange symmetry, effective dimension $d$ and short-range interaction potential, the general approach~\eref{eq:sm:ESpartialshift} together with the $\mathcal{N}$ prescription~\eref{eq:sm:ESfullshiftN}, \eref{eq:sm:DEinftymatched} is still a fully valid and meaningful ansatz.
Matching within this ansatz the next-to-leading order contribution in the expansion~\eref{eq:sm:ESN0shift} of the RHS of Eq.~\eref{eq:sm:Eshiftgeneral} with the QCE expansion [\Leref{eq:scalingNgen}] on the LHS determines the ratio
\begin{equation} \label{eq:sm:ESchimatched}
	\chi^{(N,d)}\!\left(\frac{E}{\alpha}\right) = -\frac{1}{2 c_{N-1}} g^{(N,d)}_{\pm,N-1}\!\left(\frac{E}{\alpha}\right)
\end{equation}
in terms of the coefficient functions~\eref{eq:sm:ggeneral}.
In the case of contact-interacting bosons in one dimension the latter are given by the explicit expressions~\eref{eq:sm:gfromb} and one has the exact fermionization property, implemented as
\begin{equation}
	g_{+,N-1}^{(N,d)}\!\left(\frac{E}{\alpha}\right) \xrightarrow[\alpha \to \infty]{} - 2 c_{N-1} \,,
\end{equation}
for which one reobtains the correct full shifts as $\lim_{\alpha \to \infty} \chi^{(N,d)}(E / \alpha) = \chi(0) = 1$, meaning \textit{in this case} the subscript $\infty$ can be taken literally in the sense $\Delta_\infty = \lim_{\alpha \to \infty} \Delta_\alpha$.
In the general case, where actual fermionization may be absent, the label $\Delta_\infty$ just refers to the explicit expression \eref{eq:sm:DEinftymatched}.

The analytic matching~\eref{eq:sm:ESchimatched} together with the full shifts~\eref{eq:sm:DEinftymatched} determines the partial shifts~\eref{eq:sm:ESpartialshift} as a function of $E/\alpha$ and $\mathcal{N}$.
Whenn expressing $\mathcal{N}$ as a function of $E$ it is crucial to use the unknown (already shifted) $\mathcal{N} = \mathcal{N}_\alpha(E)$ in the expression for the full shifts~\eref{eq:sm:DEinftymatched}, as demanded by the ``interaction flow consistency'' argument (see next subsection).
In order to determine the final counting function (or equivalently the shifts $\Delta_\alpha$) from known objects involves therefore a self-consistent solution:
Using the deduced analytical knowledge, Eqs.~\eref{eq:sm:ESpartialshift}, \eref{eq:sm:DEinftymatched}, \eref{eq:sm:ESchimatched}, about $\Delta_\alpha$ in the ansatz~\eref{eq:sm:Eshiftgeneral} results in a highly nontrivial algebraic equation for $\mathcal{N}_\alpha(E)$.
The most practical way to formulate the corresponding equation is by solving for $E$ with a given quantum number $\mathcal{N}$ instead of the other way around.
The process of solving can be thought of as starting with a noninteracting level of energy $E_0$ and pushing its energy until the requirement given by the matching is fulfilled.
When fixing a starting value $E_0$ the corresponding shift $\Delta_\alpha$ is determined by the equation
\begin{equation} \label{eq:sm:shiftingpracticalDelta}
	\begin{split}
		\rhoeff \Delta_\alpha &{}= \mathrm{const.} \times [\mathcal{N}_0(E_0)]^{(2/d-1)/N} \\
			&\times \chi^{(N,d)}\!\left( \frac{E_0 + \Delta_\alpha}{\alpha} \right) \,,
	\end{split}
\end{equation}
where the constant prefactor and the partial fermionization function $\chi$ are analytically given by \Eref{eq:sm:DEinftymatched} and \Eref{eq:sm:ESchimatched}, respectively.
As the root of Eq.~\eref{eq:sm:shiftingpracticalDelta}, the shifts are determined as functions
\begin{equation}
	\Delta_\alpha = \Delta_\alpha(d,N,V_\eff,E_0,\alpha) \,.
\end{equation}
of $E_0$ and $\alpha$ as well as the fixed system parameters $d,N,V_\eff$.
Equivalently one can solve for the partial fermionization $\chi$ [see also \Leref{eq:xequ}] as a function of $E_0$ and $\alpha$ by finding the root $x$ of
\begin{equation} \label{eq:sm:shiftingpracticalx}
	x = \chi^{(N,d)}\!\left( \frac{E_0 + x \, \Delta^{(N,d)}_\infty(E_0,V_\eff)}{\alpha} \right) \,,
\end{equation}
giving $\chi = x$ as
\begin{equation}
	\chi = \chi(d,N,V_\eff,E_0,\alpha) \,.
\end{equation}

\subsectionQ{Interaction flow consistency---A justification of the \texorpdfstring{$\mathcal{N}$}{N}-prescription}
\label{sec:sm:consistency}
In this subsection an analytical argument is presented clarifying why the $\mathcal{N}$-dependent prescription~\eref{eq:sm:ESfullshiftN} is favorable to an energy dependent one~\eref{eq:sm:ESfullshiftE}.
The argument is based on an infinitesimal version of the shifting method.
Instead of applying the shifting to noninteracting counting functions $\mathcal{N}_0$, Eq.~\eref{eq:sm:LCFnonint}, or equivalently individual MB levels, in order to approximately reproduce the interacting case, the starting point is here the case of finite arbitrary coupling strength $\alpha$.
One could think of the situation with the coupling set to $\alpha$ as an unperturbed system, while an infinitesimal increase $\rmd \alpha$ in the interaction can be regarded as a small perturbation.
The attempt is then to implement this perturbation as an infinitesimal version of the energy shifting method applied to $\mathcal{N}_\alpha$, expressed by
\begin{equation} \label{eq:sm:Eshiftinfansatz}
	\mathcal{N}_{\alpha + \rmd \alpha}(E) = \mathcal{N}_{\alpha}(E - \rmd E) \,.
\end{equation}
Similar to the direct finite shift~\eref{eq:sm:ESpartialshift} the ansatz for the infinitesimal shift is
\begin{equation} \label{eq:sm:EshiftinfdE}
	\rmd E = \underbrace{\mathrm{const.} \times \mathcal{N}^{(2/d-1)/N}}_{= \Delta_\infty} \rmd \chi \sim \tilde{E}^{1-d/2} \rmd \chi \,,
\end{equation}
where $\mathcal{N} = \mathcal{N}_{\alpha + \rmd \alpha}(E)$ can be regarded as the quantum number of a level to be shifted.
\begin{figure}
	\includegraphics[width=\columnwidth]{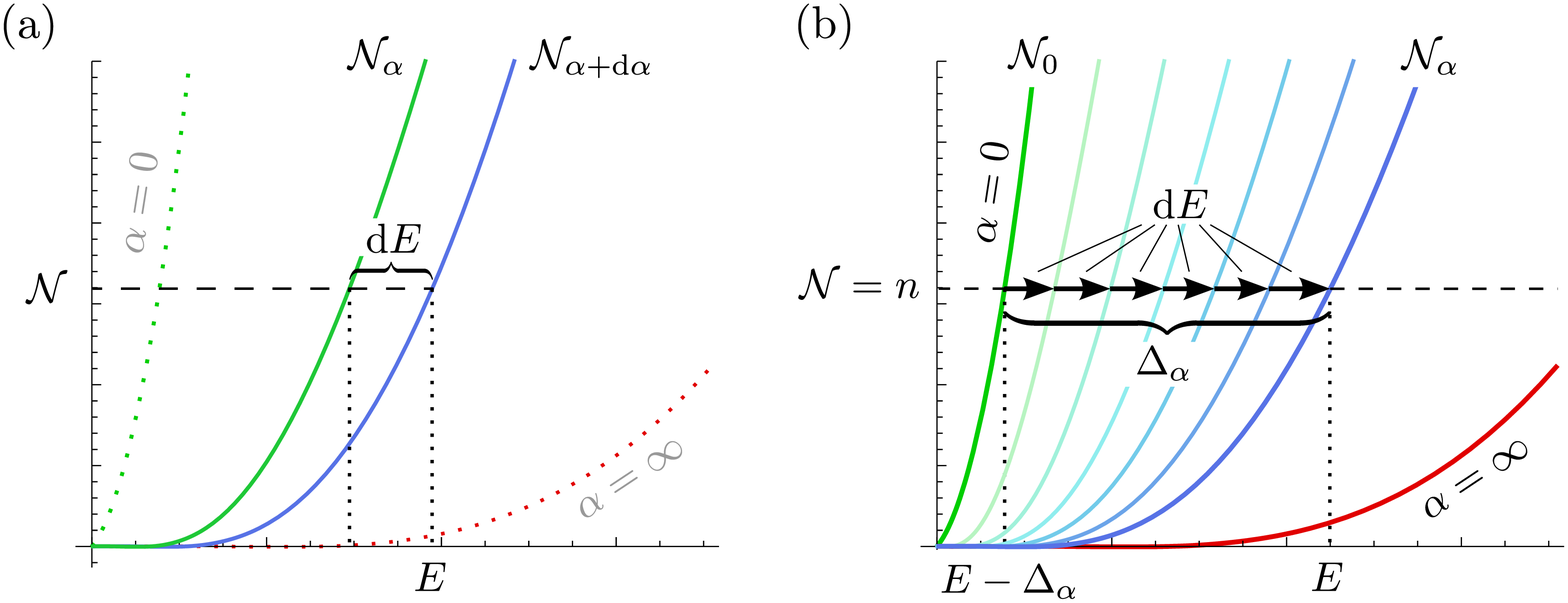}
	\caption{\label{fig:sm:eshiftflow}
		(a) Scheme of an infinitesimal energy shifting, where $\mathcal{N}_\alpha$ is the initial point (green).
		(b) Sketch of integrated infinitesimal shifts $\rmd E$ (along $\mathcal{N} = {\rm const.}$), reproducing the finite shift $\Delta_\alpha$.
	}
\end{figure}
The situation is sketched in~\fref{fig:sm:eshiftflow}(a).
The function $\rmd \chi$ is here assumed to depend on the ratio $E / \rmd \alpha$ of energy and the (here infinitesimal \textit{additional}) coupling $\rmd \alpha$.
Again a separation into different energy scales given by $\rmd \alpha$ and $V_\eff$ is crucial.
As in the previous subsection, the shift $\rmd E$ gets then determined by matching~\eref{eq:sm:Eshiftinfansatz} term by term in an expansion in $\tilde{E}$ associated with the energy scale given by the system size.
This is here demonstrated as a first-order shift involving only the matching of terms of next-to-leading order $\mathcal{O}(\tilde{E}^{(N-1)d/2})$.
Up to this order, and expanded linearly in infinitesimal quantities, the shifted counting function, according to QCE, reads
\begin{equation} \label{eq:sm:EshiftinfNa}
	\begin{split}
		&\mathcal{N}_{\alpha}(E-\rmd E) = c_N \tilde{E}^{N d/2} \\
		&{}+ \left[ c_{N-1} + g\!\left( \frac{E}{\alpha} \right) - 2 c_{N-1} \rmd \chi
				\vphantom{\frac{4}{N d} c_{N-1} c_N^{-1} \frac{E}{\alpha} g^\prime\!\left( \frac{E}{\alpha} \right) \tilde{E}^{-\frac{d}{2}} \rmd \chi} \right. \\
		&\quad\;\;{}-\underbrace{\frac{4}{N d} c_{N-1} c_N^{-1} \frac{E}{\alpha} g^\prime\!\left( \frac{E}{\alpha} \right) \tilde{E}^{-d/2} \rmd \chi}_{\text{subdominant as $\tilde{E}\to\infty$}}
				\left.
				\vphantom{\frac{4}{N d} c_{N-1} c_N^{-1} \frac{E}{\alpha} g^\prime\!\left( \frac{E}{\alpha} \right) \tilde{E}^{-d/2} \rmd \chi}
				\right] \tilde{E}^{(N-1)d/2} \\
		&{}+ \ldots \,,
	\end{split}
\end{equation}
which has to be matched with the QCE prediction
\begin{equation} \label{eq:sm:EshiftinfNada}
	\begin{split}
		&\mathcal{N}_{\alpha + \rmd \alpha} (E) = c_N \tilde{E}^{N d/2} \\
		&\;\;\;{}+ \left[ c_{N-1} + g\!\left( \frac{E}{\alpha} \right) - g^\prime\!\left( \frac{E}{\alpha} \right) \frac{E}{\alpha^2} \rmd \alpha
		\right] \tilde{E}^{(N-1)d/2} + \ldots \,,
	\end{split}
\end{equation}
where $g(E/\alpha)$ is short hand for $g_{\pm,N-1}^{(N,d)}(E/\alpha)$.
There is a subtle issue in identifying the order of terms in $\tilde{E}$.
Since in the infinitesimal shift the energy scale regarded distinctly from $\tilde{E}$ should be $E/ \rmd \alpha$ instead of $E/\alpha$, the latter could be associated with the volume scale by $E/ \alpha = \tilde{E} / \tilde{\alpha}$ with a scaled dimensionless parameter $\tilde{\alpha} = \rhoeff \alpha$.
However, $\alpha$ could be considered as a third energy scale, fixed as system parameter of the starting point system, independent of both, the (additional) interaction $\rmd \alpha$ and the system size $V_\eff$.
The corresponding behaviour $\tilde{\alpha} \to \infty$ as $V_\eff \to \infty$ keeps the ratio $E/\alpha = \tilde{E} / \tilde{\alpha}$ finite when considering the regime of large volume that underlies the expansion in dominant powers of $\tilde{E}$.
Moreover, even if $\alpha$ is not considered as fixed parameter but rather scaling with $V_\eff$, including the otherwise subdominant terms in Eqs.~\eref{eq:sm:EshiftinfNa} and \eref{eq:sm:EshiftinfNada} does not affect the discussion and leads to the same result, as will be shown in the following.
The only restriction is then that those terms do not become predominant, which would require that $g(\epsilon) \sim \epsilon^{d/2}$ when $\epsilon \to \infty$, in clear contradiction to the assumption that the effect of interaction vanishes as $\alpha \to 0$, expressed as $g(E/\alpha) \to 0$.
Recognizing that
\begin{equation}
	\frac{\rmd}{\rmd \alpha} g\!\left(\frac{E}{\alpha}\right) = g^\prime\!\left(\frac{E}{\alpha}\right) \left( - \frac{E}{\alpha^2} + \frac{1}{\alpha}\frac{\rmd E}{\rmd \alpha} \right) \,,
\end{equation}
and using Eq.~\eref{eq:sm:EshiftinfdE} allows to refine
\begin{equation}
	\begin{split}
		&\frac{E}{\alpha^2} g^\prime\!\left(\frac{E}{\alpha}\right) \rmd \alpha = - \rmd g\!\left(\frac{E}{\alpha}\right) \\
		&\qquad\qquad{}+ \underbrace{\frac{4}{N d} c_{N-1} c_N^{-1} \frac{E}{\alpha} g^\prime\!\left( \frac{E}{\alpha} \right) \tilde{E}^{-d/2} \rmd \chi}_{\text{subdominant as $\tilde{E}\to\infty$}}
			+ \ldots \,.
	\end{split}
\end{equation}
The matching then directly leads to the infinitesimal shift
\begin{equation} \label{eq:sm:Eshiftinfdchida}
	\frac{\rmd \chi}{\rmd \alpha} = - \frac{1}{2 c_{N-1}} \frac{\rmd}{\rmd \alpha} g\!\left( \frac{E}{\alpha} \right) \,,
\end{equation}
implying a flow equation for $E(\alpha)$ that depends on the particular choice of $\Delta_\infty$ in Eq.~\eref{eq:sm:EshiftinfdE}.

Finally, the $\mathcal{N}$-prescription \eref{eq:sm:DEinftymatched} for $\Delta_\infty$ becomes crucial when integrating the infinitesimal shifts (see sketch in~\fref{fig:sm:eshiftflow}b) to obtain the finite shift
\begin{equation}
	\Delta_\alpha = \int_0^\alpha \rmd \alpha^\prime \; \frac{\rmd E(\alpha^\prime)}{\rmd \alpha^\prime} \,.
\end{equation}
While the energy $E(\alpha)$ of a point on the counting function, or equivalently of an individual MB level $E^{(n)}(\alpha)$, naturally changes during the integration, its quantum number $\mathcal{N} = n$ remains constant.
Thus, combining Eq.~\eref{eq:sm:EshiftinfdE} with Eq.~\eref{eq:sm:Eshiftinfdchida}, the integrated shift becomes
\begin{equation}
		\Delta_\alpha = - \underbrace{\frac{1}{2 c_{N-1}} \Delta_\infty}_{\rm const.} \times
		\left( g\!\left( \frac{E}{\alpha} \right) - 
		\right.
		\underbrace{\lim_{\epsilon \to \infty} g(\epsilon)}_{= 0}
		\left. \vphantom{g\!\left( \frac{E}{\alpha} \right)} \right) \,,
\end{equation}
which exactly coincides with the direct finite shift~\eref{eq:sm:ESpartialshift} with Eq.~\eref{eq:sm:ESchimatched}.
This feature is here referred to as interaction flow consistency:
The direct, finite version of the shifting with $\mathcal{N}$-prescription is consistent with the integrated flow when applied as infinitesimal version at all steps in between, a very special feature that is, \eg, not inherent in an energy-dependent prescription.

\subsectionQ{Generality of the method}
First, although the method was deduced based on fermionization, the matching of the ansatz~\eref{eq:sm:ESpartialshift} together with the prescription~\eref{eq:sm:ESfullshiftN} that results in Eq.~\eref{eq:sm:ESchimatched} does not rely on this peculiarity of delta-interactions in 1D.
Instead, it applies to other types of interaction, dimensionality and also statistics.
The term ``full shifts'' should then not be taken literally, whereas the final implementation of the method in form of Eqs.~\eref{eq:sm:shiftingpracticalDelta} and \eref{eq:sm:shiftingpracticalx} should still hold, even if $\alpha \to \infty$ produces infinite shifts, then reflected by $\lim_{\epsilon \to 0} \chi(\epsilon) = \infty$.

Second, since the first-order energy shift~\eref{eq:sm:ESchimatched} only depends on the two-body clusters, it does not suffer from the truncation of QCE to first order, making it exact at the smooth level.
This reduction of the universal properties of a few-body system to the solution of the isolated two-body problem in particular opens the application to generic (short-range) interaction potentials.
It also suggests an interpretation as a quantum few-body analogue of leading-order virial expansions of classical macroscopic systems, that, \eg, lead to the Van-der-Waals equation.
Both incorporate interaction effects at the two-body level and both are restricted to describing smooth features.
While the latter applies to classical grand canonical ensembles of typically macroscopic numbers of particles, our approach describes canonical ensembles of indistinguishable particles down to the regime of quantum degeneracy.


\sectionQ{Large \texorpdfstring{$N$}{N} asymptotics of partial fermionization}
\label{sec:sm:asymptotics}

\subsectionQ{The asymptotic fermionization function \texorpdfstring{$\chi^{(N,d)}(\epsilon)$}{}}

We analyse the asymptotics of the partial fermionization function $\chi^{(N,d)}(\epsilon)$ [see \Leref{eq:EshiftChi}] in the regime of $N \gg 1$. First we work on a general level that applies to arbitrary dimensionality and arbitrary short range interaction.
To first order of the shifting method this involves, due to Eq.~\eref{eq:sm:ESchimatched}, the asymptotics of the function $g_{\pm,N-1}^{(N,d)}(\epsilon)$.
Expressing the latter in terms of the interaction kernels $a^{\suppm}(s)$ via Eq.~\eref{eq:sm:ggeneral} and \eref{eq:sm:Deltaz} reduces the problem to find the large-$N$ behavior of
\begin{equation} \label{eq:sm:sAvga11}
	\epsilon^{-(N-1)d/2} \Linvs \left[ s^{-(N-1)d/2-1} a_{\subuu}^{\suppm}(s) \right](\epsilon) \,.
\end{equation}
In general, we consider here an arbitrary clustering $\mathfrak{N}=\{n_1,n_2,\ldots,n_{|\mathfrak{N}|}\}$, including also three-body and higher order clusters, possibly involving both interacting ones and noninteracting, purely symmetry-related ones.
The corresponding contribution to the spectral coefficient functions $g_{\pm,l}^{(N,d)}(\epsilon)$ involves terms of the form
\begin{equation} \label{eq:sm:sAvgf}
	\langle f(s) \rangle_\epsilon \equiv \Gamma(\mu +1) \epsilon^{-\mu} \Linvs \left[ s^{-\mu-1} f(s) \right](\epsilon) \,,
\end{equation}
where $\mu = ld/2$ with $l=|\mathfrak{N}|$ the total number of clusters in the specific contribution and $f(s)=\Pi_{n \in \mathfrak{N}} a_n(s)$ the product of interaction kernels of all nontrivial, interacting clusters.
We set $a(s) \equiv 1$ for the trivial one-body clusters as well as noninteracting cyclic clusters of more than one particle.
This definition, in view of the noninteracting amplitudes \eref{eq:sm:Zn}, is compatible with the definition \eref{eq:sm:ACscaling}

While for first-order shifting only the term \eref{eq:sm:sAvga11} with $l=N-1$ clusters is needed, the general expressions \eref{eq:sm:sAvgf} are required for shifting of order $N-l$ and higher.
A large number of particles $N$ thus implies a large number of clusters $l \simeq N$ contributing to the shifting method unless the level of approximation becomes comparable to $N$.
The results presented in the main text [see \Lfref{fig:chi}], based on only first- and second-order shifting, show that this criterion can be met easily while providing sufficient accuracy to describe average spectra down to the quantum degenerate regime.

Eqs. \eref{eq:sm:sAvga11} and \eref{eq:sm:sAvgf} can be interpreted as (inverse) thermal ``average'' in the following sense.
It transfers functions of temperature, \ie, $a_{\subuu}^{\suppm}(s)$ or in general $f(s)$, from the canonical equilibrium picture of definite temperature (represented by $s=\beta \alpha$) to the corresponding microcanonical picture of definite energy (represented by $\epsilon = E / \alpha$).
The identity
\begin{equation} \label{eq:sm:favg}
	\langle f(s) \rangle_\epsilon = \frac{
			\Linv \left[ Z_l(\beta) \beta^{-1} f(\beta \alpha) \right](E)
		}{
			\Linv \left[ Z_l(\beta) \beta^{-1} \right](E)
		}
\end{equation}
with $Z_l(\beta) \propto \beta^{-\mu} = \beta^{-ld/2}$ reveals the nature of the ensemble over which this ``average'' is taken:
$Z_l(\beta)$ can be seen as the partition function of $l$ independent, distinguishable particles.
Therefore each cluster, as far as concerns the ensemble average, acts as a single effective particle independent of the other clusters.
On the other hand, the interaction effects that add more detail to such a composite particle picture by accounting for the internal dynamics are represented by the interaction kernels in $f(s)$ and ``averaged'' over in this ensemble rather than defining it.
For instance, replacing $f(\beta \alpha) \mapsto \beta$ in Eq.~\eref{eq:sm:favg} results in an ``average'' of $\beta$ that coincides with the microcanonical definition of inverse temperature
\begin{equation}
	\bar\beta_E = \frac{\rmd}{\rmd E} \log \mathcal{N}_l(E) = \frac{d}{2} l E^{-1} 
\end{equation}
for a system of $l$ distinguishable particles at given total energy $E$, in accordance with the equipartition theorem
\begin{equation}
	\bar E_\beta = - \frac{\rmd}{\rmd \beta} \log Z_l(\beta) = \frac{d}{2} l \beta^{-1}
\end{equation}
for the average energy in the corresponding canonical ensemble at given inverse temperature $\beta$.

Considering the limit $N\to\infty$ and hence $l,\mu\to\infty$ leads then to an equivalence of the microcanonical and canonical ensembles of clusters in the sense that
\begin{equation}
	\langle f(s) \rangle_\epsilon \to f(\bar\beta_E \alpha) = f(\mu / \epsilon) \,,
\end{equation}
showing that, during the limiting process, the energy per particle $E/N$ is the quantity that should be kept fixed rather than the total energy $E$ in order to get a nontrivial result. 
Thermal fluctuations for a finite number of particles or, more precisely, clusters, lead to subdominant corrections that can be obtained as an expansion in the ``central moments'':
\begin{equation}
	\langle f(s) \rangle_\epsilon = f(\bar s) + \frac{1}{2!} f^{\prime \prime}(\bar s) \left\langle (s- \bar s)^2 \right\rangle_\epsilon + \cdots \,,
\end{equation}
where we introduced the notation
\begin{equation}
	\bar s \equiv \langle s \rangle_\epsilon = \bar\beta_E \alpha = \mu / \epsilon \,,
\end{equation}
which should be considered as a quantity of $\mathcal{O}(1)$.
From the definition \eref{eq:sm:sAvgf} the ``central moments'' can be evaluated to be
\begin{equation} \label{eq:sm:nmoment}
	\left\langle (s- \bar s)^n \right\rangle_\epsilon = \bar s^n \sum_{k=0}^n (-1)^{n-k} \binom{n}{k} \prod_{j=0}^{k-1} \left( 1 - \frac{j}{\mu} \right) \,.
\end{equation}
To analyse the dominance of higher moment corrections for large $\mu$ we write
\begin{equation} \label{eq:sm:prodjmu}
	\prod_{j=0}^{k-1} \left(1-\frac{j}{\mu}\right) = \begin{cases}
																											\sum_{l=0}^{k-1} (-\mu)^{-l} P_l(k) & k \geq 1 \\
																											1 & k = 0
	                                                 \end{cases}
\end{equation}
where $P_l(k)$ are polynomials in the natural numbers $k$ of degree $2l$, recursively defined by
\begin{equation}
	P_l(k) = \sum_{j=l}^{k-1} j P_{l-1}(j) \,, \qquad P_0(k) \equiv 1
\end{equation}
in the case $l<k$.
One can easily show that these polynomials fulfill $P_l(k) = 0$ for $l \geq k \in \mathbb{N}_0$, except for $P_0(0) = 1$.
Therefore one can lift the upper limit in the $l$-summation in Eq.~\eref{eq:sm:prodjmu} from $k-1$ to $n-1$ and reorder the sum in Eq.~\eref{eq:sm:nmoment} to get the expansion in inverse powers of $\mu$
\begin{equation}
	\left\langle (s- \bar s)^n \right\rangle_\epsilon = \bar s^n \sum_{l=0}^{n-1} (-\mu)^{-l} \sum_{k=0}^n (-1)^{n-k} \binom{n}{k} P_l(k) \,.
\end{equation}
Since the $P_l(k)$ are of degree $2l$, all terms from $\mathcal{O}(1)$ up to $\mathcal{O}(\mu^{-n/2})$ vanish, due to the identity
\begin{equation}
	\sum_{k=0}^n (-1)^{n-k} \binom{n}{k} k^{\nu} = 0 \quad \text{for} \quad \nu < n\,, \;\; \nu \in \mathbb{N}_0 \,.
\end{equation}
The higher central moments therefore are of order
\begin{equation}
	\left\langle (s- \bar s)^n \right\rangle_\epsilon = \mathcal{O}\!\left(\mu^{-\lceil n/2 \rceil}\right) \,,
\end{equation}
where $\lceil \cdot \rceil$ denotes the ceiling function.
In particular, the first-order correction $\mathcal{O}(\mu^{-1})$ only involves the ``variance'' ($n=2$) in the (inverse) thermal ``average''.
One gets
\begin{equation} \label{eq:sm:fAvgOrdersOfMu}
	\langle f(s) \rangle_\epsilon = f(\bar s) - \frac{1}{2} f^{\prime \prime}(\bar s) \bar s^2 \mu^{-1} + \mathcal{O}(\mu^{-2}) \,.
\end{equation}

We return now to the asymptotics for partial fermionization functions.
At the level of first-order energy shifting \eref{eq:sm:ESchimatched} the exact relation to the two-body interaction kernels is
\begin{equation}
	\chi^{(N,d)}(\epsilon) = - \left\langle a_{\subuu}^{\suppm}(s) \right\rangle_\epsilon 
\end{equation}
with $\mu=(N-1)d/2$, \ie, $l=N-1$ clusters in the ensemble.
For large systems $N\gg 1$, in view of Eq.~\eref{eq:sm:fAvgOrdersOfMu}, one has
\begin{equation} \label{eq:sm:chiAsympGeneral}
	\chi^{(N,d)}(N \tilde\epsilon) = - a_{\subuu}^{\suppm}\!\left( \frac{d}{2\tilde\epsilon}\right) + \mathcal{O}(N^{-1})
\end{equation}
in the fully general case and, by using Eq.~\eref{eq:sm:adelta2},
\begin{equation} \label{eq:sm:chiAsympDelta}
	\chi^{(N,d)}(N \tilde\epsilon) = 1 - \rme^{d/(2\tilde\epsilon)} \erfc\!\left(\sqrt{d/(2\tilde\epsilon)}\right) + \mathcal{O}(N^{-1})
\end{equation}
for the repulsive Dirac delta contact interaction in 1D Bose gases [see \Leref{eq:chi_infty}].
Crucial for the nontrivial limit when $N\to\infty$ is the scaling of the total energy with $N$, which we expressed by introducing $ \tilde \epsilon = \epsilon/N = E/(N \alpha) $, \ie, the energy \textit{per particle} in units of the coupling strength $\alpha$.

We note that while the asymptotic first-order partial fermionization function \eref{eq:sm:chiAsympDelta} is determined by only the dominant term in Eq.~\eref{eq:sm:fAvgOrdersOfMu}, this turns out to be not the case for energy shifting to higher order, where thermal fluctuations become non-negligible.
This is due to the fact that to higher order, a sum of more than one ``average'' of the form \eref{eq:sm:fAvgOrdersOfMu} is involved and the corresponding dominant terms $f(\bar s)$ are divergent as positive powers of $N$.
In the overall sum, however, all divergent terms cancel each other, leaving again a nontrivial limit $N\to\infty$ when $\tilde\epsilon$ is fixed.
In the case of asymptotic second-order shifting, for instance, the variance term $\sim f^{\prime\prime}(\bar s)$ survives, while higher moments are of vanishing order in $N$.

\subsectionQ{Asymptotic scaling of the full fermionization shifts \texorpdfstring{$\Delta_\infty^{(N,d)}(E_0,V_\eff)$}{}}

In the following we consider the thermodynamic limit $N,V_\eff \to \infty$ while keeping $N/V_\eff$ fixed.
Henceforth,
\begin{equation}
	X \eqTL Y \quad \text{or} \quad X \simTL Y
\end{equation}
denotes identity or approximate identity of two quantities $X$ and $Y$ in this limit, respectively.
We will show that the full energy shift $\Delta^{(N,d)}_\infty(E_0,V_\eff)$ is an (asymptotically) \textit{extensive} quantity in the sense of
\begin{equation} \label{eq:sm:DeltaInfExt}
	\Delta^{(\lambda N,d)}_\infty(\lambda E_0, \lambda V_\eff) \eqTL \lambda \Delta^{(N,d)}_\infty(E_0, V_\eff) \,.
\end{equation}
From matching fully shifted and fermionized spectra in the high temperature regime we inferred the full shift $\Delta_\infty$, Eq.~\eref{eq:sm:DEinftymatched}, in an $\mathcal{N}$-dependent prescription \eref{eq:sm:ESfullshiftN} that was motivated by interaction flow consistency (see \hyperref[sec:sm:consistency]{Appendix E, subsection 3}).
Considering $N\gg1$ in the coefficients \eref{eq:sm:clShifting}, \eref{eq:sm:zcoeffunconf} the full shift asymptotically becomes
\begin{equation}
	\rhoeff \Delta_\infty^{(N,d)}(E_0,V_\eff) \eqTL 2^{-\frac{d}{2}} \rme^{\frac{d}{2}-\frac{2}{d}} N^{1+\frac{2}{d}} [ \mathcal{N}_0(E_0) ]^{(\frac{2}{d}-1)/N} \,.
\end{equation}
In order to express all relevant quantities in a dimensionless way that additionally reflects finite (effective) particle density $n_\eff = N/V_\eff$ in the thermodynamic limit, we introduce the (intensive) unit of energy
\begin{equation} \label{eq:sm:TDLunitE}
	\mathcal{E} = \rhoeff^{-1} N^{2/d} = \frac{2 \pi \hbar^2}{m} n_\eff^{2/d} 
\end{equation}
with $\rhoeff$ given by Eq. \eref{eq:sm:rhoeff} and define the scaled energy $\esc$ and scaled full shift $\Dsc$ as the energy and full shift \textit{per particle} in units of $\mathcal{E}$, Eq.~\eref{eq:sm:TDLunitE}:
\begin{equation} \label{eq:sm:E0scDeltaInfsc}
	\esc \equiv \frac{E_0}{N \mathcal{E}} \,, \qquad \Dsc \equiv \frac{\Delta_\infty}{N \mathcal{E}} \,.
\end{equation}
For simplicity, we set $\kB = 1$ and identify the microcanonical entropy $S$ of the noninteracting Bose gas in the usual way
\begin{equation} \label{eq:sm:entropy}
	S(N, E_0, V_\eff) = \log \mathcal{N}_0(E_0) \,.
\end{equation}
The difference in using the counting function instead of the DOS becomes insignificant in the thermodynamic limit.
The only assumption needed to go on is extensivity of the entropy \eref{eq:sm:entropy} in the sense $S(\lambda N, \lambda E_0, \lambda V_\eff) \eqTL \lambda S(N,E_0,V_\eff)$, which can be inferred from extensivity of the grand potential of the non-interacting Bose gas together with equivalence of ensembles in the thermodynamic limit.
A simple dimensional analysis shows that extensivity of $S$, itself a dimensionless quantity, implies the scaling law
\begin{equation}
	S(N, E_0, V_\eff) \eqTL N s\!\left( \esc \right) \,.
\end{equation}
It follows that the scaled full shift per particle \eref{eq:sm:E0scDeltaInfsc}, expressed in terms of the microcanonical entropy per particle $s$, is
\begin{equation} \label{eq:sm:DeltaInftysc}
	\Dsc(N, E_0, V_\eff,d) \eqTL 2^{-\frac{d}{2}} \rme^{\frac{d}{2}-\frac{2}{d}} \exp\!\left[ s\!\left( \esc \right) \left(\frac{2}{d}-1\right) \right] \,,
\end{equation}
which, in view of Eqs.~\eref{eq:sm:TDLunitE} and \eref{eq:sm:E0scDeltaInfsc}, implies the extensivity \eref{eq:sm:DeltaInfExt} of $\Delta_\infty$.

From equivalence of grand canonical and microcanonical ensembles in the thermodynamic limit, formally justified by saddle point approximation, one finds
\begin{equation} \label{eq:sm:sE0}
	s\!\left( \esc \right) = \left( 1 + \frac{2}{d} \right) \left(  \Li_{d/2}[z_0(\esc)] \right)^{2/d} \esc - \log \!\left[z_0(\esc) \right]
\end{equation}
with the fugacity $z_0(\esc)$ implicitly defined by
\begin{equation} \label{eq:sm:E0z0}
	\esc = \frac{d \Li_{1+d/2}(z_0)}{2 \left[\Li_{d/2}(z_0)\right]^{1+2/d} } \,,
\end{equation}
as long as condensation effects do not play a role.
The latter can only occur for $d > 2$, when this poses the restriction
\begin{equation} \label{eq:sm:esccrit}
	\esc > \frac{d \, \zeta\!\left(1+\frac{d}{2}\right)}{ 2 \left[\zeta\!\left(\frac{d}{2}\right)\right]^{1+2/d} } \equiv \esccrit
\end{equation}
to the validity of Eqs.~\eref{eq:sm:sE0} and \eref{eq:sm:E0z0}.
The polylogarithm in Eqs.~\eref{eq:sm:sE0} and \eref{eq:sm:E0z0} is defined as $\Li_\nu (z) = \sum_{k=1}^\infty z^k/k^\nu, |z|<1$.
The implicit definition \eref{eq:sm:E0z0} can be analytically resolved in two complementary regimes.
In the low-temperature quantum regime $\esc \ll 1$ (or $\esc - \esccrit \ll 1$) the full shift reduces to (see also \cite{Comtet2007})
\begin{equation} \label{eq:sm:DsclowT}
	\Dsc\!\left( \esc, d \right) \simTL 2^{-\frac{d}{2}} \rme^{\frac{d}{2}-\frac{2}{d}} \exp\!\left[C(d) \, (\esc)^{\frac{d}{2+d}} \right]
\end{equation}
with $ C(d) = \left( \frac{4}{d^2} - 1 \right) \left[ \frac{d}{2} \zeta\!\left(1+\frac{d}{2}\right) \right]^{2/(2+d)}$.
Remarkably, Eq.~\eref{eq:sm:DsclowT} is also valid in the BEC regime $\esc < E^{\rm sc}_{0,{\rm crit}}$, while Eqs.~\eref{eq:sm:sE0} and \eref{eq:sm:E0z0} are not applicable.
This one finds by splitting off the contributions $N^{\rm cond}$,$E_0^{\rm cond}$, and $S^{\rm cond}$ to $N$, $E_0$, and $S$ that originate from the single-particle ground state while setting $z_0 = 1$ in the thermal contributions.
While the occupation of the ground state $N^{\rm cond}$ becomes macroscopic in the BEC regime, the total energy $E_0$ and total entropy $S$ are dominated by the thermal contributions in the thermodynamic limit.
Formally, the low-temperature entropy per particle in the exponent of Eq.~\eref{eq:sm:DsclowT} could also be viewed as continuation of the function implicitly defined by Eqs.~\eref{eq:sm:sE0} and \eref{eq:sm:E0z0} into the BEC regime.

In the classical regime $\esc \gg 1$, on the other hand, the entropy per particle becomes logarithmic, resulting in a simple power-law for the full shift
\begin{equation}
	\Dsc\!\left( \esc, d \right) \simTL 2^{1-d} d^{d/2-1} (\esc)^{1-d/2} \,.
\end{equation}
In the case $d=2$ (\eg, given by a harmonic trap in 1D), the scaled full shift asymptotically becomes a unique constant for all energies:
\begin{equation}
	\Dsc \eqTL \frac{1}{2} \quad \text{for} \quad d=2 \,,
\end{equation}
in accordance with the exact unscaled full shift \eref{eq:sm:ESNinftyd2}.

\begin{figure*}[ttt]
	\includegraphics[width=\textwidth]{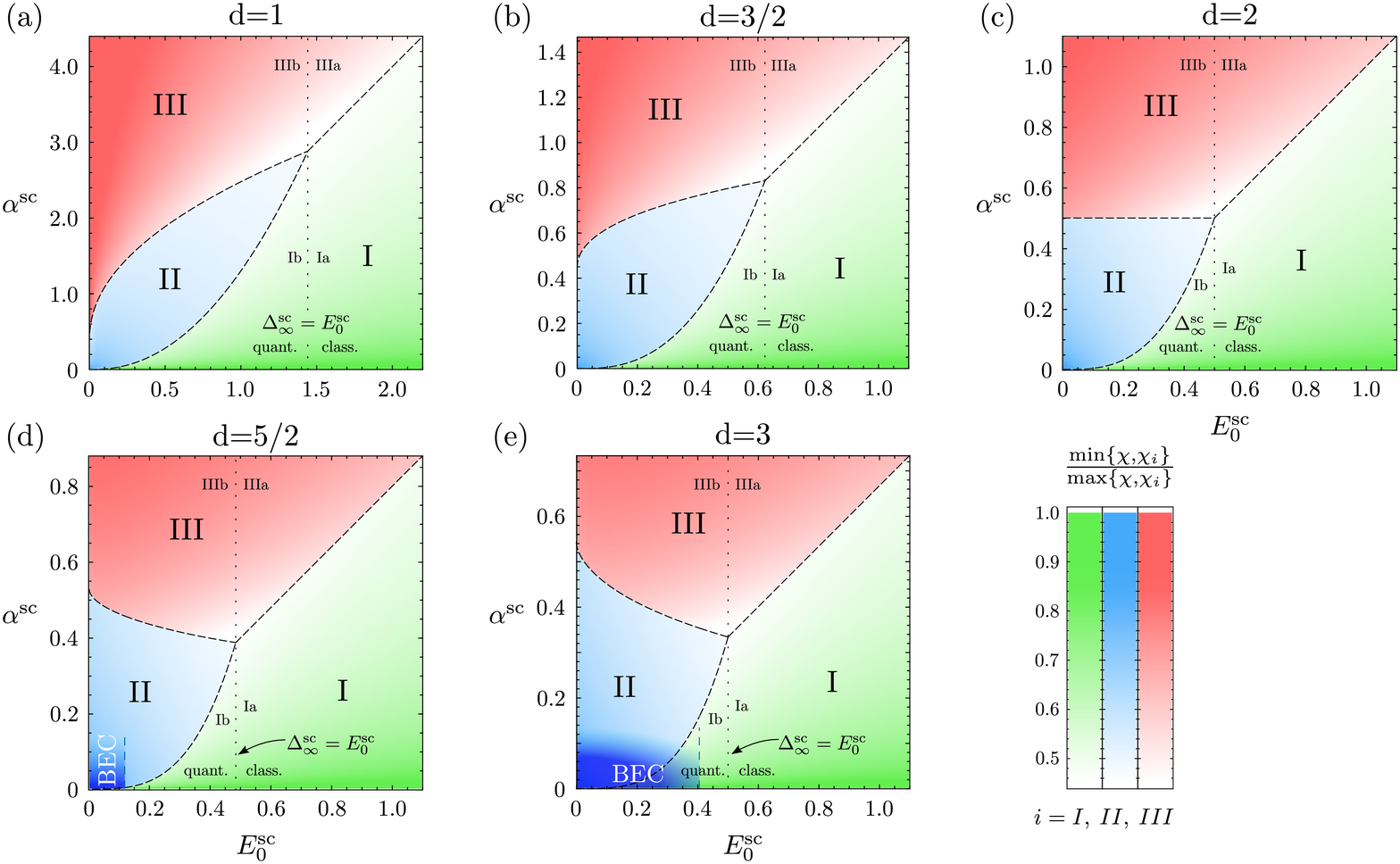}
	\caption{\label{fig:sm:regimes}
		The three regimes of partial fermionization in 1D Bose gases with contact interaction.
		I: perturbative regime, see Eq.~\eref{eq:sm:regimeI} [classical (Ia) and quantum (Ib)] (green);
		II: nonperturbative quantum regime, see Eq.~\eref{eq:sm:regimeII} (blue);
		III: fermionization regime, see Eq.~\eref{eq:sm:regimeIII} [classical (IIIa) and quantum (IIIb)] (red).
		Validity of particular regimes is encoded by lightness.
		The separation of regimes is indicated by three boundaries (dashed)
			$\asc d/2 = (\esc)^3 (\Dsc)^{-2}$ (I--II),
			$\asc d/2 = \Dsc$ (II--III), and
			$\asc d/2 = \esc$ (I--III), corresponding to Eqs.~\eref{eq:sm:regimeI}, \eref{eq:sm:regimeII}, and \eref{eq:sm:regimeIII}.
		The panels (a)--(e) show several cases of varying effective dimension:
		(a) $d=1$, \ie, periodic boundary condition and $\Vext(q) = 0$;
		(b) $d=3/2$;
		(c) $d=2$, \ie, harmonic trapping $\Vext(q) \propto q^2$;
		(d) $d=5/2$;
		(e) $d=3$, \ie, linear well, \eg, $\Vext(q) \propto |q|$.
		For $d>2$ the noninteracting Bose gas enters the BEC regime (indicated in dark blue) when $\esc \leq \esccrit$ (marked dashed blue), see Eq.~\eref{eq:sm:esccrit} .
	}
\end{figure*}

\subsectionQ{Asymptotic universality of partial fermionization \texorpdfstring{$\chi(d,\esc, \asc)$}{}}

We address the final question of the asymptotic scaling of partial fermionization as a function of the noninteracting energy $E_0$ as contrasted to the interacting energy $E = E_0 + \Delta_\alpha$ of the already shifted levels.
We express the generic asymptotic scaling \eref{eq:sm:chiAsympGeneral} of the fermionization function for arbitrary short-range interaction as
\begin{equation}
	\chi^{(N,d)}\!\left( \frac{E}{\alpha}\right) \eqTL f\!\left( \frac{2 E}{d N \alpha} \right)
\end{equation}
where, in the case of $\delta$-type contact interaction,
\begin{equation} \label{eq:sm:f}
	f(y) = 1 - \rme^{1/y} \erfc\!\left( \sqrt{1/y} \right) \,.
\end{equation}
To get $\chi$ as a function of the noninteracting energy $E_0$,
\begin{equation}
	\chi = \chi(d,N,E_0,\alpha,V_\eff) \,,
\end{equation}
one has to solve [see \Leref{eq:chi_infty}]
\begin{equation}
	\chi = f\!\left( \frac{2}{d N \alpha} \left( E_0 + \chi \Delta_\infty^{(N,d)}(E_0,V_\eff) \right) \right) \,,
\end{equation}
which, by implementing Eq.~\eref{eq:sm:DeltaInftysc}, reduces to
\begin{equation} \label{eq:sm:chiAsympImplicit}
	\chi \eqTL f\!\left( \frac{2}{d\asc} \left( \esc + \chi \Dsc(\esc,d) \right) \right) \,,
\end{equation}
where $\asc = \alpha / \mathcal{E}$ is the coupling strength in units of the intensive unit of energy $\mathcal{E}$, Eq.~\eref{eq:sm:TDLunitE}.
We arrive at the universal scaling law [\Leref{eq:universalchigen}]
\begin{equation} \label{eq:sm:chiUniversalSc}
	\chi(d,N,E_0,\alpha,V_\eff) \eqTL \chi(d,\esc,\asc) \,.
\end{equation}
The significance of Eq.~\eref{eq:sm:chiUniversalSc} is that it establishes universality in the sense that it relates smoothed spectra of interacting systems that differ in the number of particles with each other, involving a rescaling of the coupling $\alpha$ and the unshifted (noninteracting) energy $E_0$.
Remarkably, the implicit definition \eref{eq:sm:chiAsympImplicit} of the asymptotic $\chi$ even admits to write it as a function of only two parameters $2 \esc/(d \asc)$ and $2 \Dsc / (d \asc)$, which augments the universality to even unify systems that differ in effective dimension $d$, \ie, systems with all kinds of external potentials that are homogeneous functions.
This statement can be put as a relation between $\chi$ for arbitrary $d$ and the simplest case of $d=2$ via
\begin{equation}
	\chi(d,\esc,\asc) = \chi\!\left( 2, \frac{\esc}{2\Dsc}, \frac{d \asc}{4 \Dsc} \right) \,.
\end{equation}
Since it is based on the generic asymptotic scaling property of $f$, Eq.~\eref{eq:sm:chiAsympGeneral}, it applies to arbitrary short-range interaction potentials and physical dimension, as long as compatible with the QCE framework.

\sectionQ{Regimes of \texorpdfstring{$\chi$}{Chi} and explicit approximants}
\label{sec:sm:regimes}

We specify contact interaction in 1D and identify three basic regimes of the asymptotic $\chi = \chi(d,\esc,\asc)$, considering the thermodynamic limit, where the implicit definition \eref{eq:sm:chiAsympImplicit} can be resolved and shows characteristic behavior.
First, we identify the perturbative regime (I), characterized by 
\begin{equation}
	\begin{split} \label{eq:sm:regimeI}
		\asc &\ll \esc \,,\\
		\esc &\gg (\Dsc)^{2/3} (\asc)^{1/3} \,,
	\end{split}
\end{equation}
where the first term of the argument on the RHS of Eq. \eref{eq:sm:chiAsympImplicit} becomes large while the second term can be neglected, \ie, $\chi \Dsc / \esc \ll 1$, and

\begin{equation} \label{eq:sc:chiI}
	\chi \simeq f\bigl(\underbrace{2\esc/(d\asc)}_{\gg 1}\bigr) \simeq \sqrt{ 2 d \asc / (\pi \esc)} \,.
\end{equation}
This covers a ``classical perturbative regime'' (Ia) as well as a ``quantum perturbative regime'' (Ib), distinguished by either $\esc > \Dsc$ or $\esc < \Dsc$, respectively.

Second, we identify a nonperturbative quantum regime (II) by
\begin{equation}
	\begin{split} \label{eq:sm:regimeII}
	\asc &\ll \Dsc \,,\\
	\esc &\ll (\Dsc)^{2/3} (\asc)^{1/3} \,,
	\end{split}
\end{equation}
where the second term of the argument on the RHS of Eq. \eref{eq:sm:chiAsympImplicit} becomes large while the first term can be neglected, \ie, $\chi \Dsc / \esc \gg 1$, and
\begin{equation}
	\chi \simeq f\bigl(\underbrace{2 \chi \Dsc / (d\asc)}_{\gg 1}\bigr) \simeq \sqrt{2d \asc / (\pi \Dsc \chi)} \,,
\end{equation}
which simplifies to
\begin{equation}  \label{eq:sc:chiII}
	\chi \simeq  \left( \frac{2 d \asc}{\pi \Dsc} \right)^{1/3} \,.
\end{equation}

Regimes I and II are both characterized by a large argument to the function $f$ in Eq. \eref{eq:sm:chiAsympImplicit}, \ie, a low level of fermionization $\chi \ll 1$.
They are complemented by the fermionization regime (III), \ie,
\begin{equation}
	\begin{split} \label{eq:sm:regimeIII}
	\esc \ll \asc \,,\\
	\Dsc \ll \asc \,,
	\end{split}
\end{equation}
where a small argument to the function $f$ in Eq. \eref{eq:sm:chiAsympImplicit} implies
$1 - \chi \ll 1$ and one gets
\begin{equation}  \label{eq:sc:chiIII}
	\chi \simeq 1 - \sqrt{ 2 (\esc + \Dsc)/ (\pi d \asc)} \,,
\end{equation}
covering both, a ``classical fermionization regime'' (IIIa) as well as a ``quantum fermionization regime'' (IIIb).

The identification of these regimes becomes especially simple in the case $d=2$, where $\Dsc \eqTL 1/2$ is constant.
For this case, $\chi$ becomes a particularly rigid shift \wrt $\esc$ in the nonperturbative quantum regime.

\Fref{fig:sm:regimes} shows the partition of the parameter space into the regimes identified above and the validity of the respective approximations, Eqs.~\eref{eq:sc:chiI},\eref{eq:sc:chiII}, and \eref{eq:sc:chiIII}.

Finally, an overall good approximation is already obtained by a single iteration.
Using the expression for regime II, Eq.~\eref{eq:sc:chiII}, on the RHS of Eq.~\eref{eq:sm:chiAsympImplicit} gives
\begin{equation} \label{eq:sm:chiIteration}
	\chi \simeq f\!\left( \frac{2\esc}{d\asc} +  \left( \frac{4}{\pi} \right)^{1/3} \left( \frac{2\Dsc}{d\asc} \right)^{2/3} \right)
\end{equation}
with $f$ and $\Dsc(\esc)$ defined in Eqs.~\eref{eq:sm:f} and \eref{eq:sm:DeltaInftysc}, respectively.
In the full parameter range the interpolation function \eref{eq:sm:chiIteration} has a maximum relative error $\delta = |\chi_{\rm approx} -\chi|/\chi$ of $\delta < 11.7 \% $ \wrt the exact numerical solution of the implicit Eq.~\eref{eq:sm:chiAsympImplicit}.
If one applies the iteration once more, \ie, plugging Eq.~\eref{eq:sm:chiIteration} into the RHS of Eq.~\eref{eq:sm:chiAsympImplicit} results in an explicit approximation with maximum relative error $ \delta < 3.0 \%$.
The third and fourth iteration give $ \delta < 0.99 \% $ and $ \delta < 0.38 \% $.
It is worth to note that for the second and higher iterations also the relative error in $1-\chi$, \ie, $\bar{\delta} = |\chi_{\rm approx} -\chi|/(1-\chi)$, referring to the deviation from fermionization, is bound by $\bar{\delta} < 5.3 \%, 0.88 \%, 0.23 \% $, referring to the second, third and fourth iteration, respectively.
Note that it is crucial to take the approximation in regime II as initial value.
It is special in so far that it correctly produces the dominant behavior in all three regimes after just one iteration.
In contrast, taking for instance $\chi \equiv 0,1,$ or Eq.~\eref{eq:sc:chiI} as starting point results in unbounded relative errors $\delta$, even after multiple iterations.


\bibliography{qce1d_prl}

\nocite{GarciaGarcia2017,Bogomolny2014}

\end{document}